\documentclass[12pt,a4paper,twoside]{article}
\makeatletter\if@twocolumn\PassOptionsToPackage{switch}{lineno}\else\fi\makeatother

\usepackage{amsfonts,amssymb,amsbsy,latexsym,amsmath,tabulary,graphicx,times,xcolor}
\usepackage[T1]{fontenc}
\usepackage{url,multirow,morefloats,floatflt,cancel,tfrupee,multicol}
\makeatletter

\AtBeginDocument{\@ifpackageloaded{textcomp}{}{\usepackage{textcomp}}}
\makeatother
\usepackage{colortbl}
\usepackage{xcolor}
\usepackage{pifont}
\usepackage[nointegrals]{wasysym}
\usepackage[superscript,biblabel]{cite}
\makeatletter

\definecolor{mygreen}{rgb}{0, 0.6, 0}
\newcommand{\blue}[1]{{\color{blue}#1}}
\newcommand{\red}[1]{{\color{red}#1}}
\usepackage{authblk}
\usepackage{comment}

\@ifundefined{etal}{\def\etal{\textit{et~al}}}{}

\AtBeginDocument{
\expandafter\ifx\csname eqalign\endcsname\relax
\def\eqalign#1{\null\vcenter{\def\\{\cr}\openup\jot\m@th
  \ialign{\strut$\displaystyle{##}$\hfil&$\displaystyle{{}##}$\hfil
      \crcr#1\crcr}}\,}
\fi
}

\@ifundefined{tsGraphicsScaleX}{\gdef\tsGraphicsScaleX{1}}{}
\@ifundefined{tsGraphicsScaleY}{\gdef\tsGraphicsScaleY{.9}}{}
\def\checkGraphicsWidth{\ifdim\Gin@nat@width>\linewidth
	\tsGraphicsScaleX\linewidth\else\Gin@nat@width\fi}

\def\checkGraphicsHeight{\ifdim\Gin@nat@height>.9\textheight
	\tsGraphicsScaleY\textheight\else\Gin@nat@height\fi}

\def\fixFloatSize#1{}%\@ifundefined{processdelayedfloats}{\setbox0=\hbox{\includegraphics{#1}}\ifnum\wd0<\columnwidth\relax\renewenvironment{figure*}{\begin{figure}}{\end{figure}}\fi}{}}
\let\ts@includegraphics\includegraphics

\def\inlinegraphic[#1]#2{{\edef\@tempa{#1}\edef\baseline@shift{\ifx\@tempa\@empty0\else#1\fi}\edef\tempZ{\the\numexpr(\numexpr(\baseline@shift*\f@size/100))}\protect\raisebox{\tempZ pt}{\ts@includegraphics{#2}}}}

\AtBeginDocument{\def\includegraphics{\@ifnextchar[{\ts@includegraphics}{\ts@includegraphics[width=\checkGraphicsWidth,height=\checkGraphicsHeight,keepaspectratio]}}}

\DeclareMathAlphabet{\mathpzc}{OT1}{pzc}{m}{it}

\def\URL#1#2{\@ifundefined{href}{#2}{\href{#1}{#2}}}

\edef\fntEncoding{\f@encoding}

\makeatother

\newif\ifmultipleabstract\multipleabstractfalse%
\usepackage[hmargin=1.8cm,vmargin=2.2cm]{geometry}%,headsep=0.3in,footskip=0.49in
\makeatletter
\def\author#1{\gdef\@author{\hskip-\dimexpr(\tabcolsep)\hskip1pt\parbox{\dimexpr\textwidth-1pt}{\centering #1}}}

\let\@articletype\@empty \def\articletype#1{\gdef\@articletype{{\fontsize{14}{16}\selectfont #1}}}

\usepackage{fancyhdr}

\fancypagestyle{headings}{\fancyhf{}\fancyhead[RE]{\MakeTextUppercase{\textbf{\RunningAuthor}}}\fancyhead[LE]{\textbf{\thepage}}\fancyhead[LO]{\MakeTextUppercase{\textbf{\RunningHead}}}\fancyhead[RO]{\textbf{\thepage}}}
\fancypagestyle{plain}{\fancyhf{}\fancyfoot[C]{\textbf{\thepage}}}

\AtBeginDocument{\usepackage{lastpage}}
\def\title#1{%
  \gdef\@title{%
    \ifx\@articletype\@empty\else\@articletype~\\\fi%
     #1}%
}

\usepackage{abstract}
\def\abstractname{\textbf{Abstract}}
\renewenvironment{onecolabstract}
{\vspace*{-.4pc}\trivlist\item[]\leftskip1pt\noindent\selectfont\hfill\abstractname\hfill\mbox{\null}\par\ignorespaces}{\endtrivlist}

\def\NormalBaseline{\def\baselinestretch{1.1}}

\usepackage{textcase}
\usepackage[explicit]{titlesec}
\titleformat{\section}[block]{\NormalBaseline\boldmath\bfseries}
{\thesection.}
{6pt}
{#1}
[]
\titleformat{\subsection}[hang]{\NormalBaseline\filright\itshape}
{\thesubsection.}
{6pt}
{#1}
[]
\titleformat{\subsubsection}[runin]{\NormalBaseline\filright\itshape}
{\hspace{16pt}\thesubsubsection}
{6pt}
{#1}
[]
\titleformat{\paragraph}[runin]{\NormalBaseline}
{\theparagraph}
{6pt}
{#1}
[]
\titleformat{\subparagraph}[runin]{\NormalBaseline}
{\thesubparagraph}
{6pt}
{#1}
[]

\titlespacing{\section}{0pt}{1.5\baselineskip}{.2\baselineskip}  
\titlespacing{\subsection}{0pt}{1.5\baselineskip}{.2\baselineskip}  
\titlespacing{\subsubsection}{0pt}{1.5\baselineskip}{.2\baselineskip}  
\titlespacing{\paragraph}{0pt}{.5\baselineskip}{10pt}  
\titlespacing{\subparagraph}{0pt}{.5\baselineskip}{10pt}

\usepackage{caption}
\DeclareCaptionLabelFormat{figlabel}{Figure #2}
\date{}
\makeatother

\usepackage{endfloat}
\usepackage{float}

\begin{document}

\title{Fluid-driven traveling waves in soft robots}
\def\RunningHead{
Fluid-driven traveling waves in soft robots
}
\def\RunningAuthor{salem \etal}

\author{Lior Salem, Amir D. Gat and Yizhar Or
\thanks{Lior Salem, Amir D. Gat and Yizhar Or are with Faculty of Mechanical Engineering and TASP- Technion Autonomous Systems and Robotics Program.
       Technion - Israel Institute of Technology,
Technion City, Haifa, Israel 3200003
        {\tt\small liorsal@campus.technion.ac.il}}%
}

\maketitle

%%%%%%%%%%%%%%%%%%%%%%%%%%%%%%%%%%%%%%%%%%%%%%%%%%%%%%%%%%%%%%%%%%%%%%%%%%
%TC:ignore
{\begin{onecolabstract}
Many marine creatures, gastropods, and earthworms generate continuous traveling waves in their bodies for locomotion within marine environments, complex surfaces, and inside narrow gaps. In this work, we study theoretically and experimentally the use of embedded pneumatic networks as a mechanism to mimic nature and generate bi-directional traveling waves in soft robots. We apply long-wave approximation to theoretically calculate the required distribution of pneumatic network and inlet pressure oscillations needed to create desired moving wave patterns.  We then fabricate soft robots with internal pneumatic network geometry based on these analytical results. The experimental results agree well with our model and demonstrate the propagation of moving waves in soft robots, along with locomotion capabilities. The presented results allow fabricating soft robots capable of continuous moving waves via the common approach of embedded pneumatic networks and requiring only two input controls.

\def\keywordstitle{Keywords}
\end{onecolabstract}}
%TC:endignore 
%%%%%%%%%%%%%%%%%%%%%%%%%%%%%%%%%%%%%%%%%%%%%%%%%%%%%%%%%%%%%%%%%%%%%%%%%%
\begin{multicols}{2}

\section{Introduction}
Many invertebrate creatures, such as gastropods, cuttlefish, and flagellated microorganisms, generate traveling waves in their bodies for locomotion, exploiting their flexibility in order to move in various unpredictable environments. Soft-robots, which have the ability to deform their flexible geometry, are inspired by these natural organisms and aim to mimic their capabilities \cite{Trivedi2008,Kim2013,Polygerinos2017}. Consequently, soft robots have diverse techniques for locomotion that suit their surroundings, such as legged locomotion \cite{Shepherd.2011,Tolley2014,umedachi2013highly,mao2014gait}, crawling \cite{kamamichi2006snake,Boxerbaum2012,rafsanjani2018kirigami}, rolling \cite{onal2017soft,lin2011goqbot}, swimming  \cite{Feng2019BodyWave,marchese2014autonomous,mojarrad1997biomimetic} etc.

Previous studies examined, both theoretically and experimentally, various approaches for the actuation of traveling waves for locomotion of soft robots \cite{dobrolyubov1986mechanism,chirikjian1995kinematics,chen2003analysis,chen2007design}. Among the suggested methods, are electro-active actuation  \cite{Stalbaum2017, Jafferis2013, Poole2012} and surface wave robots \cite{setter2012low,setter2014propulsion, suzumori2006new}. Yet, the most commonly used approach to actuate soft robots is by pressurizing embedded fluidic channels, creating large amplitude waves.

Multiple inlets may actuate such robots, where each elastic segment is controlled separately to create a wave-like motion \cite{Onal2013, luo2015slithering, branyan2017soft}. However, reducing the number of inlets may result in a more simple structure containing less hardware such as tubes, valves, and controllers. This approach may be applicable in confined spaces and in cases where lightweight and minimal implementation is crucial, such as in space missions \cite{lin2013soft}.

In order to realize a reduced number of controlled inlets for creating traveling wave motion, several works presented designs that consist of three inflated chambers which actuated sequentially with a phase difference. By transferring between three different geometries, the authors were able to create semi-continuous and approximate traveling wave motion in soft robots. More specifically, Takayama et al. \cite{Takayama2015} presented a soft robot composed of three bonded and twisted inlet tubes. The inlet pressure was alternating between the different tubes to generate helical rotation, which was utilized for locomotion through a pipe. Ozaki et al. \cite{ozaki2011novel} developed a three-chamber actuator that was wrapped around an endoscope and actuated sequentially to create traveling waves on the surface which exploited for self-propulsion. A design of a soft robot which generates transverse waves, presented by Watanabe et al. \cite{Watanabe2017} consists of a sophisticated arrangement of three internal rubber tubes, limited-strain sleeve, and fabric layers that restrain radial and axial expansion. Similar to previous cases, the three tubes are pressurized sequentially, generating transverse traveling wave motion and crawl on the surface. Another relevant work by Qi et al. \cite{qi2020novel} exhibited a 3D printed modular robot, designed using finite element software, that uses four types of similar modules that differ only by their internal air pathways. Thus, modules can be assembled serially and controlled by four inlets to form approximated traveling wave motion regardless of the number of modules. Moreover, several studies leveraged viscosity and fluid dynamics to minimize the number of controlled inlets for generating propulsion\cite{Vasios2019, Futran2018,Glozman2010,Lim2008}.

The current work aims to demonstrate the use of an embedded pneumatic network, a common actuation method in soft robots, to mimic nature and generate continuous bi-directional traveling waves while using only two inlets. To achieve this, we derive a theoretical model and apply standard and simple fabrication techniques to experimentally demonstrate this approach. We then analyze the generated waves to calibrate the inlets and compare our analytic and experimental results.

\section{Analysis - Traveling waves based on two pneumatic network }
Spatially distributed fluid channels embedded in the elastic body in addition to time-dependent pressure inlets are combined to create traveling wave motion. This model-based design reduces the number of inlets to two. This section presents the derivation of the design concept.

\begin{figure}
    \centering
   %trim={<left> <lower> <right> <upper>}

    %\includegraphics[clip, trim=0cm 2.5cm 6cm 0cm,width=1\linewidth]{figures/FigDeformations2.pdf}
    %\includegraphics[width=0.8\textwidth,natwidth=610,natheight=642]{tiger.pdf}
    \includegraphics[width=1\linewidth]{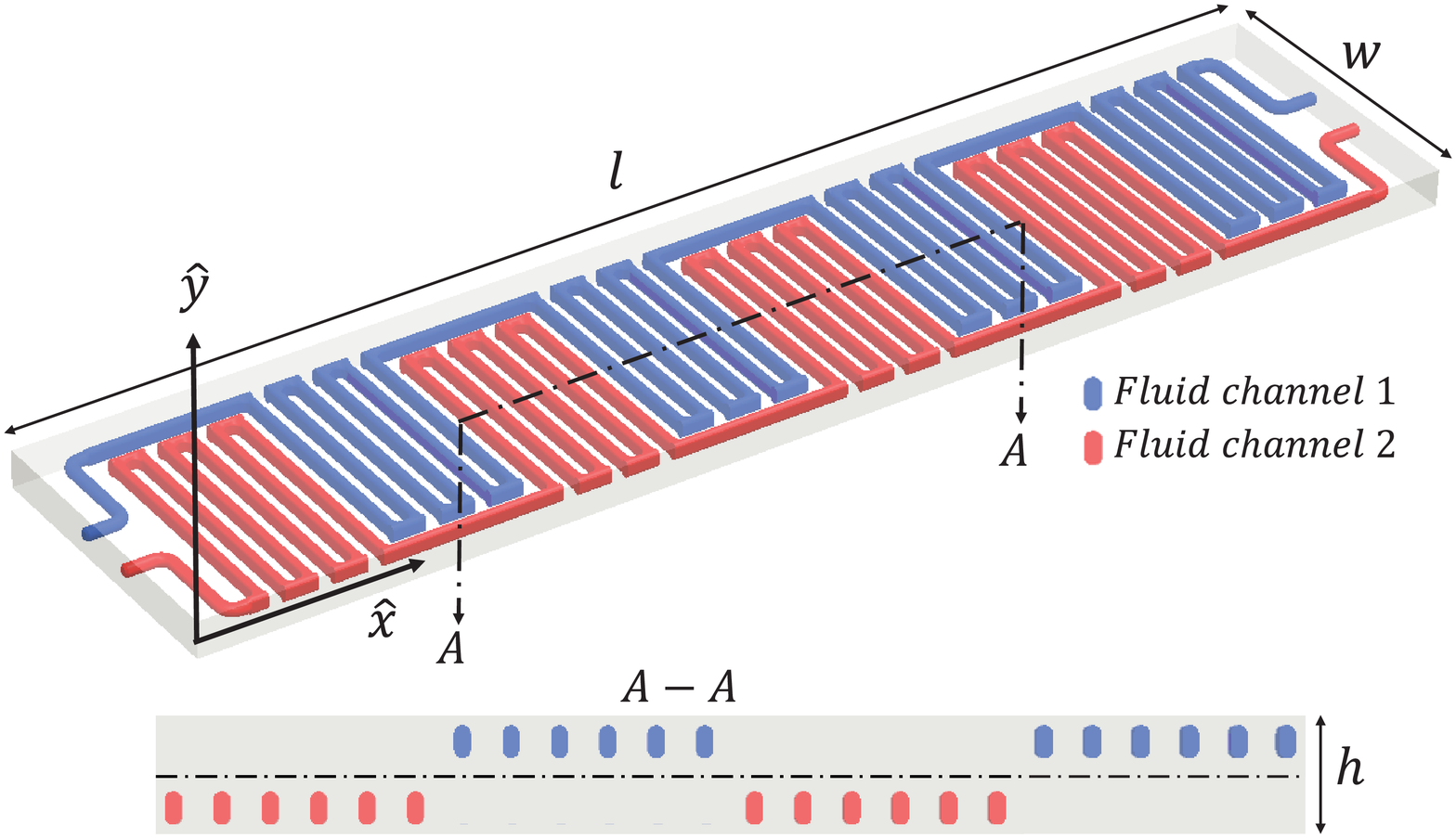}
    \caption[Embedded fluid-filled channel network]{Illustration of an elastic structure with two sets of embedded fluid-filled channel network (top), and its side view (bottom). In this specific illustrated case, \blue{Channel network 1} is always above neutral surface expressed by positive $\phi_1 (x)$ values, and \red{Channel network 2} is always below neutral surface represented by negative $\phi_2 (x)$ values.}
   \label{fig:EFN}
\end{figure}

We model the robot as a rectangular cross-section elastic beam with height $h$, width $w$ and length $l$, with embedded parallel channel network. The calculation of the beam's deformation induced by pressure can be realized from the modified Euler-Bernoulli equation as reported in previous works, \cite{Matia2017,Gamus2017}  
\begin{multline}\label{modified-euler-bernoulli}
        \frac{\partial^2 }{\partial x^2}\left[EI\left(\frac{\partial^2 y}{\partial x^2}+\phi(x) p(x,t) \frac{1}{E}\frac{\partial \psi}{\partial (p/E)}\right)\right]+\\ +c\frac{\partial y}{\partial t}+\rho_s A_s\frac{\partial^2 y}{\partial t^2}=q,
\end{multline}
where  $x$ is an actuator-spatial longitudinal coordinate, $y=y(x,t)$ is lateral deformation, $E$  is Young's modulus,  $I$ is second moment of the beam cross-section, $\phi(x)$ is channel density distribution (number of channels per unit length), $p(x,t)$ is fluid pressure in channels, $\partial \psi / \partial (p/E)$ is beam slope change due to a single channel at pressure $p$, $c$ is structural damping, $\rho_s$ is solid density, $A_s$ is cross-section area and $q(x,t)$ is distributed external load. See Figure \ref{fig:EFN}.

In order to create the desired deformation pattern, our design aims to calculate the channel distribution, which is combination of two indexed channels, $\phi_1(x), \phi_2(x)$,  and corresponding time-dependent pressure inlets $p_1(t), p_2(t)$. Applying equation [\ref{modified-euler-bernoulli}], with the assumption of negligible inertia and viscosity and no external loads yields a relation between deformation and two sets of channels and inputs,

\begin{equation}\label{d2ydx2}
-\frac{\partial^2 y}{\partial x^2}=\frac{\partial \psi}{\partial (p/E)}\left(\phi_1(x)p_1(t)+\phi_2(x)p_2(t)\right).
\end{equation}
The desired deformation pattern of a single wavelength and a single frequency traveling wave is described as,
\begin{equation}\label{TWform2}
        y(x,t)=A\cos\left(\omega t-\frac{2\pi x}{\lambda} \right),
\end{equation}
where $A$ is the wave amplitude, $\omega$ is its frequency, and $\lambda$ is the spatial wavelength. Another equivalent form to represent  this traveling wave is

\begin{equation}\label{TWform1}
\begin{split}
 y(x,t)=A \Bigl(\cos\left(\frac{2\pi x}{\lambda} \right)\cos(\omega t)+ \\
\sin\left(\frac{2\pi x}{\lambda} \right)\sin(\omega t) \Bigr)
\end{split}
\end{equation}

%\begin{multline}\label{TWform1}
%           y(x,t)=A\left[\cos\left(\frac{2\pi x}{\lambda} \right)\cos(\omega t)+\\ +\sin\left(\frac{2\pi x}{\lambda} \right)\sin(\omega t) \right].
%\end{multline}

Substituting relation [\ref{TWform1}] into equation [\ref{d2ydx2}] , without loss of generality, we may assume that, 

\begin{subequations}\label{phi_p_combinations}
\begin{equation}
\begin{aligned}
  \frac{\partial \psi}{\partial (p/E)}\phi_1(x)&p_1(t)=\\
  &A\left(\frac{2\pi}{\lambda}\right)^2\cos\left({\frac{2 \pi x}{\lambda}}\right)\cos{(\omega t}),
  \end{aligned}
  \end{equation}
  \begin{equation}
  \begin{aligned}
    \frac{\partial \psi}{\partial (p/E)}\phi_2(x)&p_2(t)=\\&A\left(\frac{2\pi}{\lambda}\right)^2\sin\left({\frac{2 \pi x}{\lambda}}\right)\sin{(\omega t)}.
\end{aligned}
\end{equation}
\end{subequations}

For the inlet pressures

\begin{equation}\label{p1p2_straight}
   p_1(t)=p_0 \cos{(\omega t)} \quad\text{,}\quad  p_2(t)=p_0 \sin{(\omega t)}
\end{equation}

the required channel densities are obtained as

  \begin{equation}\label{phi1phi2}
  \begin{aligned}
    &\phi_1(x)=\phi_0\cos(2\pi x / \lambda)\\& \phi_2(x)=\phi_0 \sin(2\pi x / \lambda)
\end{aligned}
\end{equation}

and $p_0=A\kappa ^2/(\phi_0\partial \psi / \partial (p/E) )$, where $\kappa=2\pi/\lambda$ is the wavenumber.

\begin{figure}
    \centering
   %trim={<left> <lower> <right> <upper>}
    %\includegraphics[clip, trim=0cm 2.5cm 6cm 0cm,width=1\linewidth]{figures/FigDeformations2.pdf}
    \includegraphics[clip, trim=0cm 1cm 2cm 0cm,width=1\linewidth]{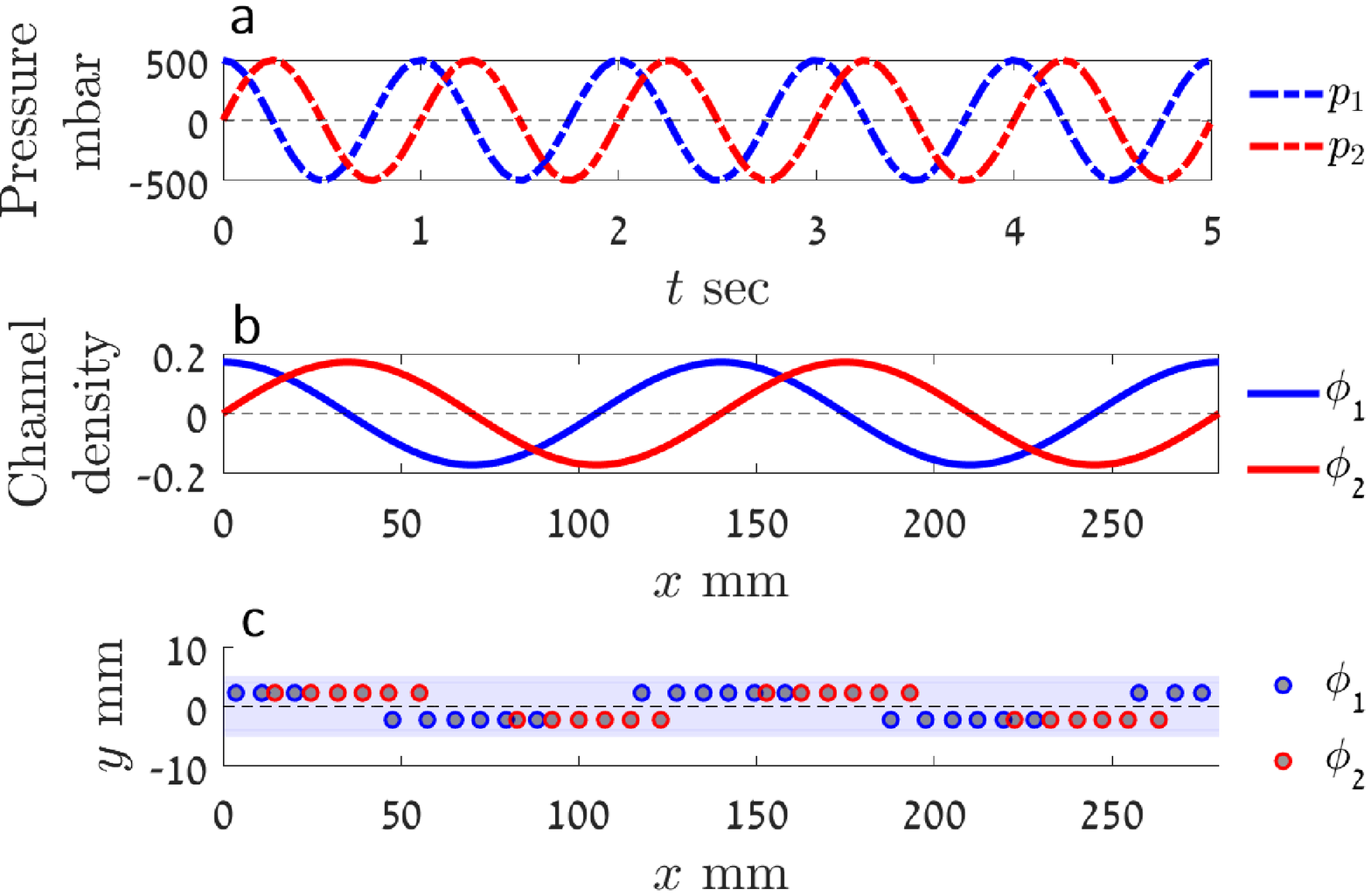}
    \caption[Theoretical straight robot configuration.]{Theoretical straight robot configuration. \textbf{(a)} Inlet pressures of channels 1 and 2 are marked by dashed \textcolor{blue}{blue} and \red{red} lines, respectively. \textbf{(b)} Channel densities of sets 1 and 2 are marked by \textcolor{blue}{blue} and \red{red} lines, respectively. \textbf{(c)} Robot's side view presents initial straight shape and channel locations. Positive values of channel density function $\phi_i(x)$ denote channels that are located above the neutral surface, whereas negative values are located below it.}
   \label{fig:StraightRobotInputs}
\end{figure}

Positive values of channel density function $\phi(x)$ denote channels that are located above the beam's neutral surface, where negative values are located below it. See Figure~\ref{fig:EFN}. Equation [\ref{phi1phi2}] represents a continuous channel distribution function and needs to be translated to discrete locations. The center of the $k^{th}$ channel, $x_k$, is calculated by\cite{Matia2015} \begin{equation}\label{discrete_locations}
  \int_{0}^{x_k} |\phi (x)|\,dx=k-\frac{1}{2},
\end{equation}
  where $k$ is a natural number. Figure \ref{fig:StraightRobotInputs} describes the theoretical channel density function and the inlet pressure of two channels. It also illustrates the locations of the discrete channels, which are located above the neutral surface when the corresponding density function, $\phi_1(x)$ or $\phi_2(x)$  is positive and are located below the neutral surface if $\phi_1(x)$ or $\phi_2(x)$ is negative.

This simple design comprises only two sets of channels and is fabricated easily due to its initial planar shape. We notice that the inlet pressures $p_1(t)=p_0 \cos{(\omega t)} $, $ p_2(t)=p_0 \sin{(\omega t)}$ dictate negative gauge pressures which are limited in magnitude and, in turn, cause a limited deformation. Note that working with liquids in a negative pressure range may initiate phase change. Furthermore, we assume a linear relation between the pressure, $p$, and the slope change angle, $\psi$, namely that $\partial \psi / \partial (p/E)$ is constant. Experimentally, this assumption is valid for a certain range of positive gauge pressures but is not satisfied by negative values (see in Supplementary Information (SI) Figure \ref{fig:NPslopes}). Thus for certain cases, it might be beneficial to avoid negative gauge pressures.

In order to obtain non-negative inlet pressures, we may modify them as follows,
  \begin{equation}\label{p1p2_positive}
  \begin{aligned}
    &p_1(t)=p_0\left( \cos{(\omega t)+1}\right)\\& p_2(t)=p_0\left( \sin{(\omega t)+1}\right)
\end{aligned}
\end{equation}

However, the additional terms multiplied by the channel density function, according to equation [\ref{d2ydx2}], form an undesired curvature invariant with respect to time. To overcome this problem, we may use a third auxiliary channel which is needed to cancel these terms with constant pressure $p_3$, where the channel density $\phi_3(x)$ may be expressed as,

\begin{equation}\label{ThirdChannel}
\begin{aligned}
   &\frac{\partial \psi}{\partial (p/E)}\phi_3(x)p_3=\\&-A\left(\frac{2\pi}{\lambda}\right)^2\left[\sin\left({\frac{2 \pi x}{\lambda}}\right)+\cos\left({\frac{2 \pi x}{\lambda}}\right)\right].
\end{aligned}
\end{equation}

\begin{figure}[h]
    \centering
   %trim={<left> <lower> <right> <upper>}
    \includegraphics[clip, trim=0cm 1cm 2cm 0cm,width=1\linewidth]{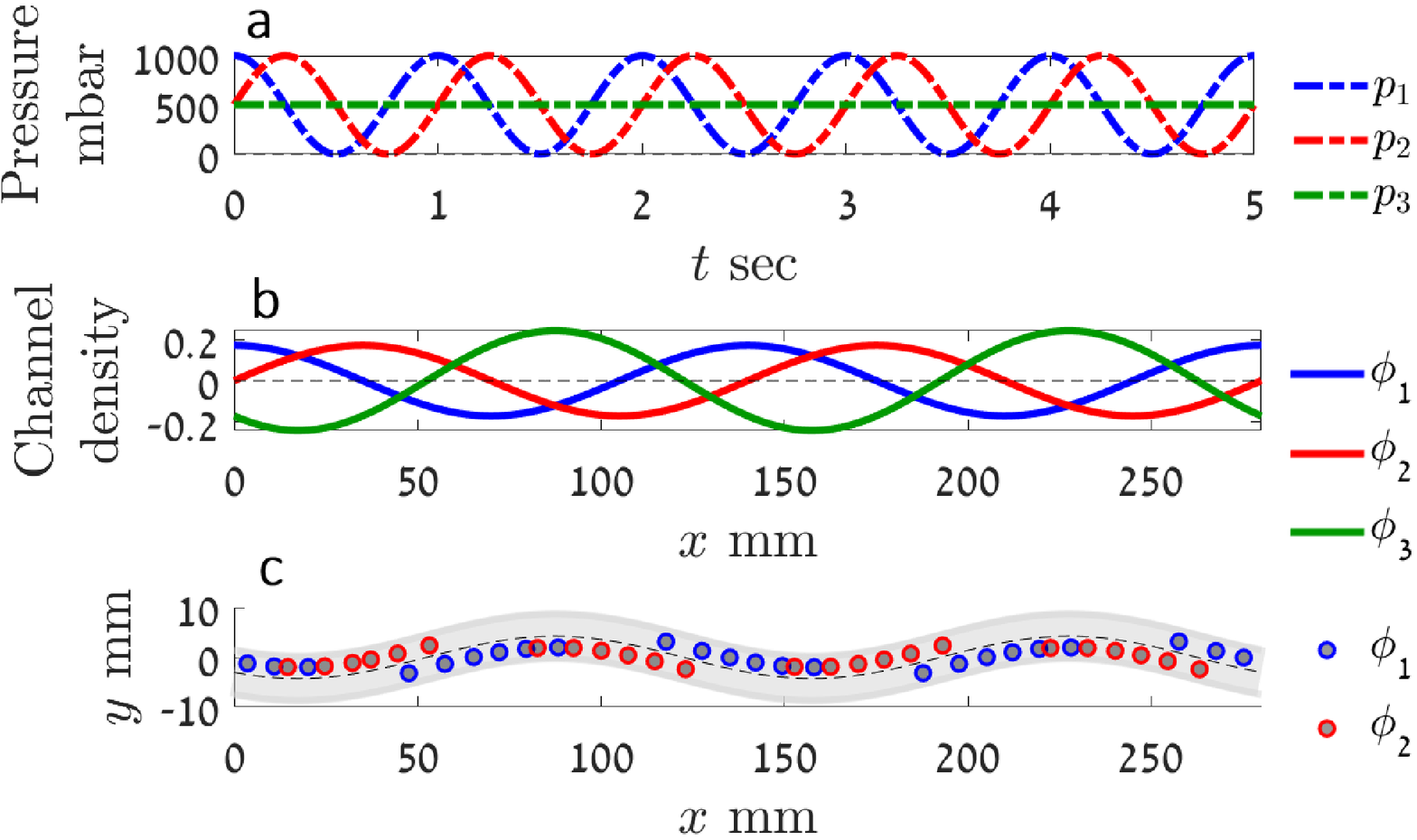}
 \caption[Theoretical pre-curved robot configuration.]{Theoretical pre-curved robot configuration. \textbf{(a)} Inlet pressures of channels 1, 2, and auxiliary channel 3 are marked by dashed \textcolor{blue}{blue}, \red{red} and \textcolor{mygreen}{green} lines, respectively. \textbf{(b)} Channel densities of sets 1, 2 and auxiliary channel 3 are marked by \textcolor{blue}{blue}, \red{red} and \textcolor{mygreen}{green} lines, respectively. \textbf{(c)} Robot's side view presents pre-curved shape $y_0$ and channel locations. Positive values of channel density function $\phi_i(x)$ denote channels located above the neutral surface, whereas negative values mean location below it.}
   \label{fig:PreCurvedInputs}
\end{figure}

This approach enables to change the wave amplitude of the beam only by adjusting the unbounded applied inlet pressure of all three channels. However, multiple channels increase the possibility of the intersection of the channels' cross-sections, which increases the fabrication complexity, and reduces the beam's homogeneity. Hence, to avoid an additional third channel, as well as the third controlled inlet, we may design the beam in a pre-curved shape instead of the auxiliary channel's constant deformation contribution (see Figure \ref{fig:PreCurvedInputs}). A drawback of this approach is the pre-determined amplitude without the ability to tune it as before. From equations [\ref{d2ydx2}] and  [\ref{ThirdChannel}], we calculate the required shape of the pre-curved beam, $y_0(x)$, as follows
\begin{equation}\label{PreCurvedShape}
    y_0(x)=-\sqrt{2}A\sin\left({\frac{2 \pi x}{\lambda}+\frac{\pi}{4}}\right).
\end{equation}

\section{Fabrication and experimental setup}

\subsection*{Fabrication considerations and methods}
\begin{figure}
    \centering
   %trim={<left> <lower> <right> <upper>}

    \includegraphics[clip, trim=5.5cm 0cm 7cm 0cm,width=1\linewidth]{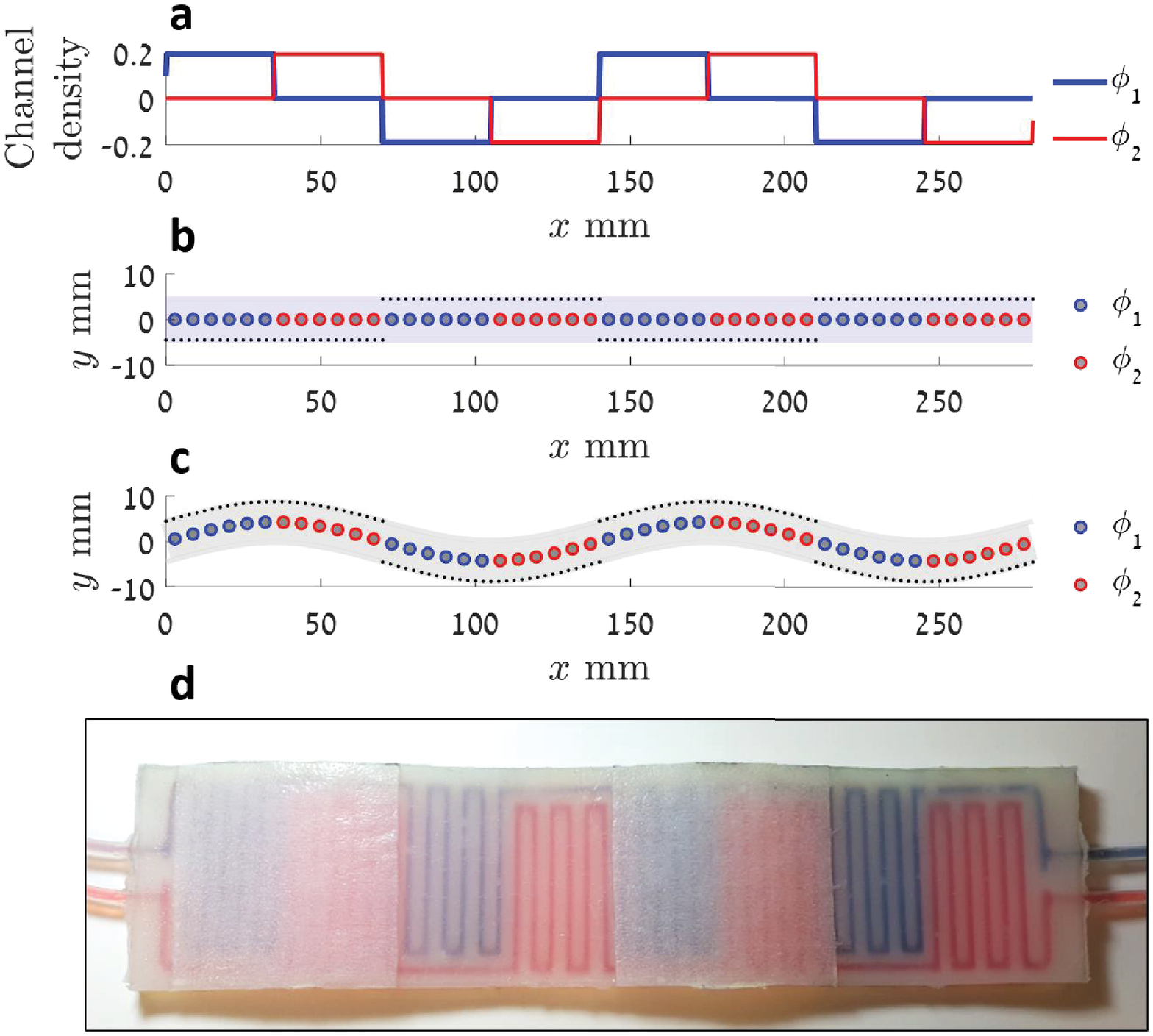}

    \caption[Robots' design.]{Robots' design. \textbf{(a)} Channel density functions $\phi_1(x)$ and $\phi_2(x)$.  \textbf{(b)} Straight robot discrete channels locations with added limited-strain layers marked by dashed lines.  \textbf{(c)} Pre-curved robot discrete channels locations with added limited-strain layers marked by dashed lines. The length of the limited-strain layers is described by equation [\ref{FiberLength}]. \textbf{(d)} Robot's top-view displays two sets of channels and top surfaces' limited-strain layers.}
   \label{fig:Design}
\end{figure}

In the previous section, equation [\ref{phi1phi2}] presents a continuous channel distribution function and its mapping into its discrete locations using equation [\ref{discrete_locations}]. Subject to channel distributions, the number of discrete channels, and cross-section geometry, two channels may overlap. Also, the varying signs of $\phi_1 (x)$ and $\phi_2 (x)$ place the channels both above and below the neutral surface, which increases fabrication complexity and yields a thicker robot.  According to theoretical model guidelines, we placed the channels with the constraints of non-intersections and a minimal gap of 1 channel diameter between the channels. For each $x$ coordinate, only one channel is located. Limited-strain layers are attached to the top and bottom surfaces, alternately determining the neutral surface and curvature direction. Limited-strain layers replace the need to locate the channels in the distance from the center of structures' cross-section, thus allowing the channels to lie on the center. We found that minor changes in channel locations and the use of limited-strain layers yield a thinner structure and simplify fabrication with little effect on traveling wave motion relative to discrete theoretical design. The design is illustrated in Figure \ref{fig:Design} both for initially straight and pre-curved robots. The soft robot fabrication process is presented in Figure \ref{fig:FabExp} and detailed in SI.

\subsection*{Experimental setup}
\begin{figure*}
    \centering
  % trim={<left> <lower> <right> <upper>}
    \includegraphics[clip, trim=0cm 20cm 0cm 2cm,width=1\linewidth]{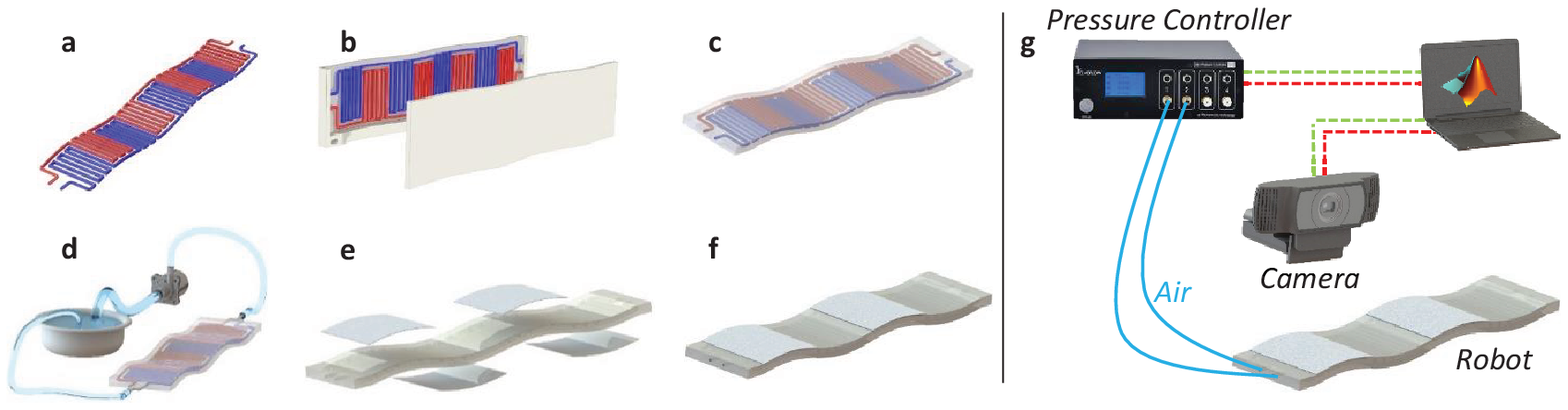}
    \caption[Fabrication process and experimental setup.]{Fabrication process and experimental setup. \textbf{(a)} 3D printed PVA channels cores. \textbf{(b)} PLA mold for silicone casting around a channels core. \textbf{(c)} Elastic structure with embedded PVA core. \textbf{(d)} PVA core dissolving process using circulating water pump. \textbf{(e)} Limited-strain layers glued to surfaces. \textbf{(f)} Fabricated robot. \textbf{(g)} Experimental setup includes simultaneous trigger on Pressure controller which is connected to two robot's inlets and frame grab via HD camera.}
   \label{fig:FabExp}
\end{figure*}

The robot was placed on a flat surface, and a pressure controller (Elveflow OB1) was connected to its inlets. A camera (Logitech C920, HD, 30fps ) was placed perpendicular to the deflection plane and captured frames simultaneously to actuation. In order to extract deformation signal, we used MATLAB\textsuperscript{\textregistered} image processing toolbox to adjust frames prior to region classification. Utilizing high contrast between the robot's and background color, we were able to apply a simple algorithm on the bi-modal histogram to determine threshold value and to extract the robot's shape in time $y(x,t)$. The experimental setup is illustrated in Figure \ref{fig:FabExp}(g).

\section{Experimental results}

This section presents the experimental results of wave propagation in two types of robots: 1. Initially straight robot (see Figure \ref{fig:Design}(b)) having two pressure inlets, varying from negative to positive gauge pressure, in the range from -675 mbar to 530 mbar. 2. Pre-curved robot (see Figure \ref{fig:Design}(c)) having two inlets subject to positive gauge pressure exclusively, from 300 mbar to 1200 mbar. First, we present the method of measurements, data analysis, and identification to quantify the wave propagation. We display a traveling wave ratio (TWR) map for various inlet pressures for both robots, indicating the preferable inlets to create traveling wave motion.

\subsection*{Experimental identification}
We now briefly review an approach to fit a curve to measured data using tempo-spatial decomposition in order to quantify the propagation wave's quality. This method is well established in  \cite{VERED2021107515, bucher2019experimental, bucher2004estimating,setter2014propulsion}, and extensively detailed in Supplementary Information (SI). The outcome of the process may help tuning the real system's controlled inlets in order to achieve an effective traveling wave. 

Once the deformation measurements $y(x,t)$ have been extracted from images, the deformation signal can be decomposed in time and space \cite{setter2014propulsion} and takes the form of a truncated Fourier series which describes the wave as a sum of several one-dimensional waves that have different frequencies and wavelengths. The first harmonic associated with the robot's designed wave-number and the excitation frequency has greater amplitude, in order of magnitude, than the other modes. While it is possible to analyze each one of the decomposed one-dimensional waves, all other modes have minor contribution to the propagation. Hence, we are interested only in the first harmonic which is further analyzed. 

The first harmonic, composed of a single frequency and wavelength, can be written as a combination of two traveling waves propagating in opposite directions,
\begin{equation}\label{complex-amplitudes}
    y(x,t)=\Re\left(U_+e^{i (\omega t-\kappa x)}+U_-e^{i (\omega t+\kappa x)}\right),
\end{equation}
where $U_+$ and $U_-$ are the complexes forward and backward wave's amplitudes, respectively, which can be obtained from the tempo-spatial decomposition (see SI) and determine the behavior of a single wave, including its magnitude and direction.

The traveling wave ratio (TWR) is a scalar parameter that quantifies a single wave propagation characteristics which is possible due to the modal decomposition. We can define the TWR as the ratio between the standing wave and the mixed-wave amplitude \cite{bucher2004estimating,VERED2021107515},

\begin{equation}\label{TWR_definition}
    TWR=1-\frac{||U_+|-|U_-||}{|U_+|+|U_-|}.
\end{equation}
This bounded ratio, $0 \leq TWR \leq 1$, defines the propagation of a wave, where $TWR=0$ denotes pure traveling wave, whereas $TWR=1$ denotes pure standing wave. Any other value between $0$ to $1$ expresses a mixed traveling and standing wave. Moreover, the wave's direction can be obtained by the sign of $|U_+|-|U_-|$, positive or negative, corresponds to forward or backward direction, respectively. 

The described identification procedure gives rise to map the propagation behavior of the robot for each set of inputs, and helps tuning the input parameters in order to create traveling wave.

\subsection*{Straight robot with two inlets of negative to positive gauge pressure range}

A straight beam shape design consists of two sets of channels with time phase inputs and spatial phase distribution as presented above. The gradient $\gamma^+\triangleq \partial \psi/\partial (p/E)^+$ is nearly constant for a certain positive range of pressure. For the negative gauge pressure range, the gradient is also approximately constant. However, it has a moderately smaller slope where $ \gamma^+\approx1.27\gamma^-$ as shown in SI Figure \ref{fig:NPslopes}(a). In order to create the same deformation amplitudes for both positive and negative gauge pressure inlets, we need to compensate for the lower slope by applying larger absolute pressure as shown in SI Figure \ref{fig:NPslopes}(b) satisfying $p_{0}^-\gamma^+=p_{0}^+\gamma^-$ . Thus, inlet pressures are calculated by

\begin{equation}\label{PressureCompensation}
   \begin{aligned}
    p_1(t)=\min[p_{0}^+\cos{(\omega t)},p_{0}^-\cos{(\omega t)}],\\      
    p_2(t)=\min[p_{0}^+\sin{(\omega t)},p_{0}^-\sin{(\omega t)}].
  \end{aligned}
\end{equation}

Once controlled inlets for negative gauge pressure are calibrated, we denote by $\hat{p}_i$ the nominal inlet pressure where $\hat{p}_i=p_{0}^+$ and $i=1,2$ is the inlet number. Generally, two sinusoidal inlets have three controlled parameters: $\hat{p}_1,\hat{p}_2$, and phase shift $\theta$, given by
\begin{equation}\label{p1p2_straight_general_wave}
   p_1(t)=\hat{p}_1 \sin{(\omega t+\theta)} \quad\text{,}\quad  p_2(t)=\hat{p}_2 \sin{(\omega t)}.
\end{equation}
Aiming to create a pure traveling wave pattern, we may set $\hat{p}_1=\hat{p}_2$  and phase shift of $\theta=\pi/2$ which essentially yields equation [\ref{p1p2_straight}]. However, dealing with an actual system, we expect inaccuracies caused by fabrication methods, design compromises, and unmodeled phenomena. Hence, tuning the three controlled inlets $\hat{p}_1,\hat{p}_2$, and $\theta$ may improve the quality of the traveling wave.

\begin{figure}
    \centering
    
    %trim={<left> <lower> <right> <upper>}
    %\includegraphics[clip, trim=5cm 0cm 4.5cm 0cm,width=1\linewidth]{ExpResults_StraightRobot.eps}
 
    \includegraphics[width=0.5\linewidth]{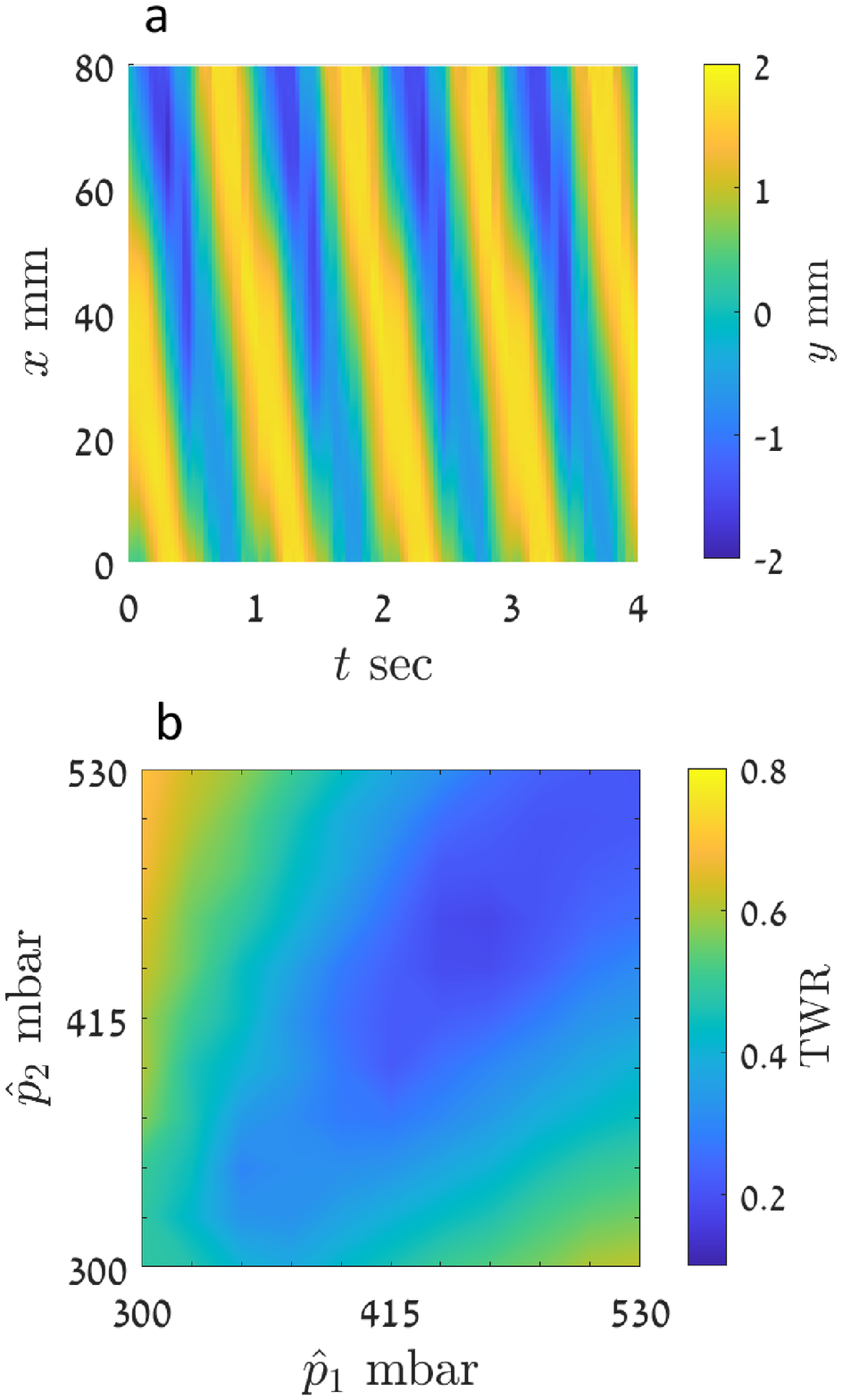}
    \caption[Experimental results of measurements and data analysis for the straight robot.]{Experimental results of measurements and data analysis for the straight robot. \textbf{(a)} A typical image of  wave propagation. Color map represents measured deflection, $y(x,t)$,  demonstrating traveling wave pattern in space and time for $\hat{p}_1=530$ mbar, $\hat{p}_2=530$ mbar, and $\theta=\pi/2$. \textbf{(b)} Color map represents low (cool) and high (warm) values of TWR for different values of $\hat{p}_i$ introduced at channels $i=1,2$, and phase shift of $\theta=\pi/2$. Low TWR values which are located approximately along the principal diagonal (theoretical diagonal at $\hat{p}_1=\hat{p}_2$) represent propagating waves. Tick marks on axes denote the 11x11 grid of values for which the measurements were undertaken.}
   \label{fig:IdentificationStraight}
\end{figure}

In order to tune the controlled parameters, they were tested experimentally at the ranges between $\hat{p}_1,\hat{p}_2=300$ to $530$ mbar with 11 equally spaced values and phase shift $\theta$ from $0$ to $\pi$ with 11 equally spaced values. The measured tempo-spatial deformation under each inlet set was analyzed and quantified via TWR value, as explained in the previous section. Figure \ref{fig:IdentificationStraight}(a) illustrates the robot's traveling wave pattern in space and time at $\hat{p}_1,\hat{p}_2=530$ and $\theta=\pi/2$. Figure \ref{fig:IdentificationStraight}(b) presents TWR map as function of controlled parameters where, for brevity, a smooth 2D surface was generated from 11x11 grid points at $\theta=\pi/2$  using \emph{contourf} command in MATLAB\textsuperscript{\textregistered}. Theoretically, we expect to get minimum values of TWR, i.e. best traveling waves, along the principal diagonal line where $\hat{p}_1=\hat{p}_2$  according to equation [\ref{p1p2_straight}]. Along the diagonal, waves are close to pure traveling waves, and their amplitudes increase as $\hat{p}_1$ and $\hat{p}_2$ become larger. We can observe the expected diagonal yet with a slight deviation from the exact symmetry line $\hat{p}_1=\hat{p}_2$. This deviation is associated with model inaccuracies and small asymmetries in the structure and two-channel sets.

\subsection*{Pre-curved robot having two inlets subject to positive gauge pressure}

Using only positive gauge pressure simplifies actuation, avoids cavitation issues, and also can create larger amplitudes. Based on the previous theoretical section, we created a pre-curved elastic robot, where its initial shape is according to equation [\ref{PreCurvedShape}] with wavelength of $\lambda=l/2$ and amplitude of $A\approx2$ mm. Figure \ref{fig:Design} illustrates its design.

Similar to the previous section, equation [\ref{phi1phi2}] still represents the channels' densities along the $x$ coordinate, and the same considerations regarding the practical fabrication of embedded channel network are valid. Equation [\ref{p1p2_positive}] represents the applied inlet pressure of each channel in the case of exclusively positive gauge pressure, and in this case, $\gamma= \partial \psi/\partial (p/E)$ is nearly constant for the whole range of motion.

\begin{figure*}
    \centering
        %trim={<left> <lower> <right> <upper>}
   \includegraphics[clip, trim=1cm 0cm 1cm 0cm,width=1\linewidth]{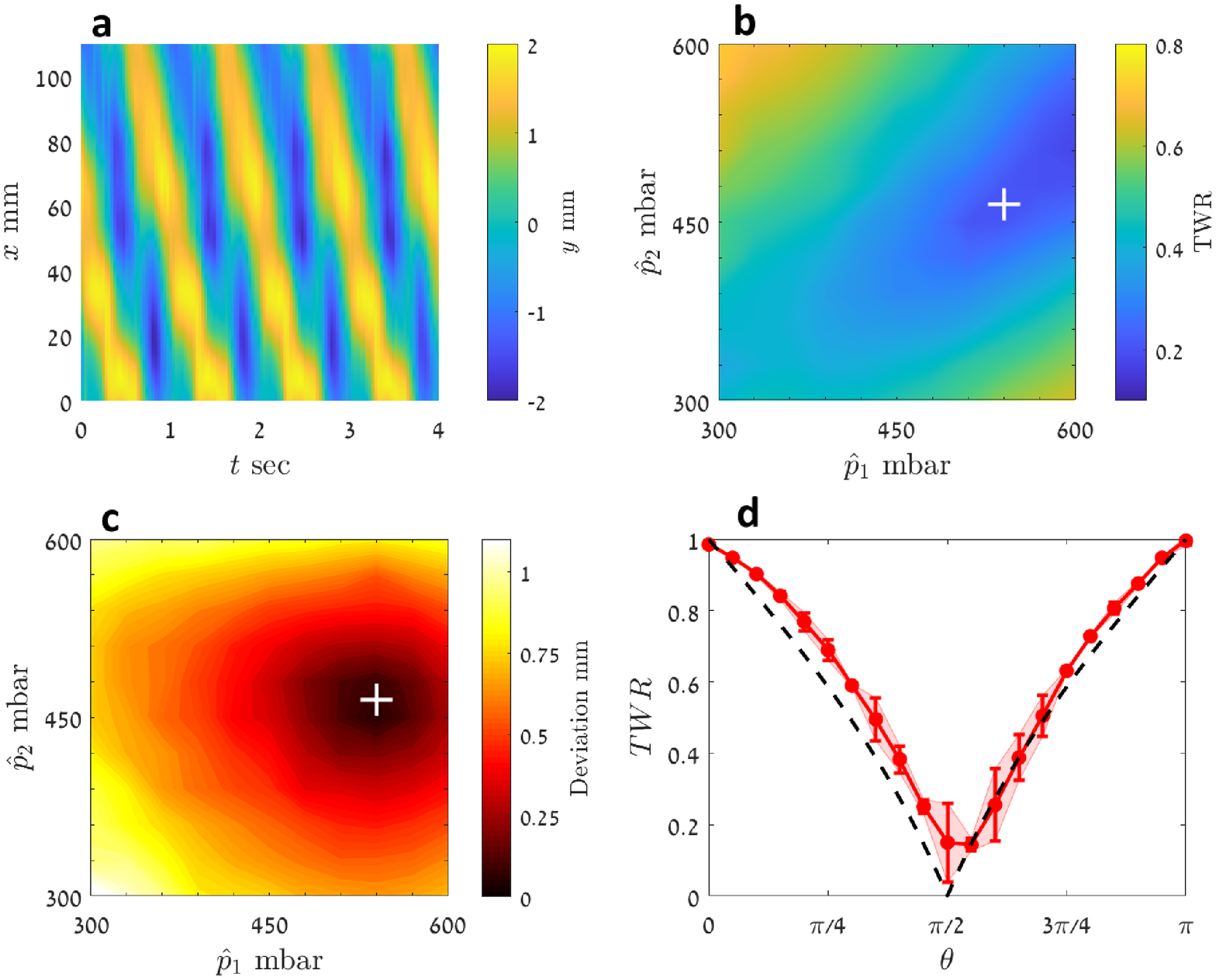}

    \caption[Experimental results of measurements and data analysis for the pre-curved robot.]{Experimental results of measurements and data analysis for the pre-curved robot. \textbf{(a)} A typical image of wave propagation. Color map represents measured deflection, $y(x,t)$,  demonstrating propagating wave pattern in space and time for $\hat{p}_1=540$ mbar, $\hat{p}_2=480$ mbar, and $\theta=\pi/2$. \textbf{(b)} TWR map for different values of $\hat{p}_i$ introduced at channels $i=1,2$. Color map represents low (cool) and high (warm) values of TWR. Low TWR values are located approximately along the principal diagonal line. Working point is marked by "$+$". \textbf{(c)} Lowest deviation amplitude from the desired wave is nearly zero at the working point pressure, marked by "$+$". Tick marks on axes denote the 11x11 grid of values for which the measurements were undertaken. \textbf{(d)} The effect of temporal phase shift, $\theta$, on the measured TWR in a pre-curved robot for $\hat{p}_1=540$ mbar and $\hat{p}_1=480$ mbar, is marked by red dots. Measured TWR function versus $\theta$ attains minimum at approximately $\theta=\pi/2$. The corresponding analytic curve of TWR as a function of $\theta$ is marked by a dashed black line.}
   \label{fig:IdentificationPreCurved}
\end{figure*}

As previously, we aim to tune the three controlled parameters  $\hat{p}_1,\hat{p}_2$, and $\theta$, in order to achieve pure traveling wave motion as close as possible. The inlets are given by,

\begin{equation}\label{p1p2_precurved_general_wave}
\begin{aligned}
   &p_1(t)=\hat{p}_1 (\sin(\omega t+\theta)+1)\\& p_2(t)=\hat{p}_2 (\sin(\omega t)+1). 
\end{aligned}
\end{equation}
Due to the symmetry, theoretically we may set  $\hat{p}_1=\hat{p}_2$  and phase shift of $\theta=\pi/2$, which yields the pure traveling wave inlets as stated in equation [\ref{p1p2_positive}]. However, due to the mentioned inaccuracies, we wish to tune the parameters. We experimentally tested the robot with 11 equally spaced values of the pressure at the inlets in the range from $\hat{p}_1,\hat{p}_2=300$ to $600$ mbar and 11 equally spaced values of the phase shift, $\theta$, in the range from $0$ to $\pi$. Each experiment was analyzed and the TWR of each wave was obtained. Figure \ref{fig:IdentificationPreCurved}(a) illustrates the propagating wave in space and time at $\hat{p}_1=540$ mbar, $\hat{p}_2=480$ mbar, and $\theta=\pi/2$. Figure \ref{fig:IdentificationPreCurved}(b) presents TWR map as a function of controlled parameters where, for brevity, smooth 2D surface was generated from 11x11 grid points at $\theta=\pi/2$ using \emph{contourf} command in MATLAB\textsuperscript{\textregistered}. The area in which the lowest value of TWR (cold colors) is displayed, is located approximately along the principal diagonal line, where $\hat{p}_1=\hat{p}_2$. Unlike the case of the straight robot, now the ideal wave's amplitude is pre-determined and cannot be altered. Thus, ideal performance is expected at $ \hat{p}_1=\hat{p}_2=A\kappa ^2/(\phi_0\gamma )$. Due to the mentioned inaccuracies, the experimental working point, marked by "$+$" on the TWR map approximately at $\hat{p}_1=540$ mbar, $\hat{p}_2=480$ mbar, is not positioned exactly on $ \hat{p}_1=\hat{p}_2$ line. Any applied pressures lower than the working point pressures will be insufficient to equalize the amplitude opposite to the pre-curved shape. Similarly, applied pressures greater than the working point increase the amplitude of the opposite side beyond the pre-curved robot designed amplitude. In these cases, propagating modes, if there are any, will not oscillate around the desired $y=0$ axis but around a deviation function $d(x)\approx D\sin\left(\kappa  x+\vartheta\right)$. The ideal (desired) case is $D \approx 0$, where the robot oscillates about $y=0$, i.e. the inlet pressures are roughly equal to working point pressures. In the case where inlet pressures are zeroed, $d(x)$ takes the form of the initially curved robot. Figure \ref{fig:IdentificationPreCurved}\textbf{(c)} presents the color map of the deviation amplitude $D$ for varying $\hat{p}_1,\hat{p}_2$. The experimental working point is marked by "$+$" representing the working point at $\hat{p}_1=540$ mbar, $\hat{p}_2=480$ mbar where the deviation get its lowest value, $D\approx 0$.

Previous TWR maps (Figures \ref{fig:IdentificationStraight} and \ref{fig:IdentificationPreCurved}(a)-(c)) were calculated for various ranges of $\hat{p}_1,\hat{p}_2$ at the fixed phase shift of $\theta=\pi/2$, as stated theoretically in pressure inlets' equations [\ref{p1p2_straight}] and [\ref{p1p2_positive}]. Figure \ref{fig:IdentificationPreCurved}(d), which exhibits experimentally the influence of the temporal phase shift $\theta$ on the TWR, displays a minimum value of TWR at approximately $\theta=\pi/2$ for fixed values of $\hat{p}_1=540$ mbar and $\hat{p}_2=480$ mbar. An analytic expression for TWR as a function of $\theta$ for one-dimensional wave may be expressed as,

\begin{equation}\label{analytic_TWR_vs_phase}
   TWR(\theta)=1-\frac{\left| \sqrt{1+\sin \theta}-\sqrt{1-\sin \theta}\right|}{ \sqrt{1+\sin \theta}+\sqrt{1-\sin \theta}}.
\end{equation}
This analytic expression is overlaid in Figure \ref{fig:IdentificationPreCurved}(d) by a dashed line, showing a very good agreement with the experimentally measured value.

\section*{Traveling-wave robot locomotion}

This section presents the underlying kinematics of locomotion of the traveling wave soft robots on the ground, and shows the experimental demonstration of soft robots in several scenarios.

\subsection*{Kinematic analysis}
\begin{figure}
    \centering
    \includegraphics[width=1\linewidth]{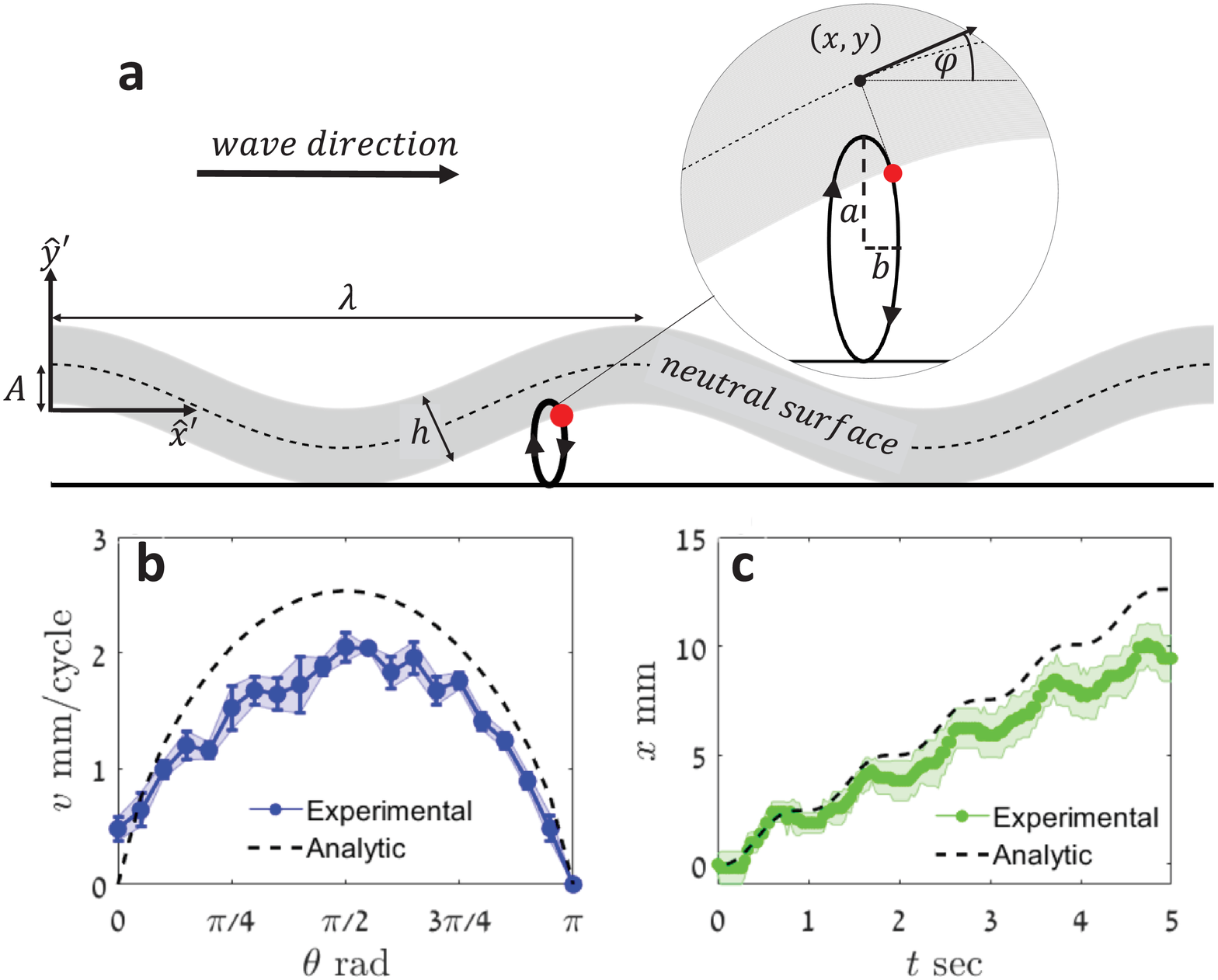}
    \caption[Traveling wave robot locomotion illustration.]{Traveling wave robot locomotion illustration. The neutral surface oscillates as an one-dimensional wave, where $A$ is the wave amplitude, $\lambda$ is the wavelength, and $h$ is robot's height. A path of the material point on a surface contacting the ground is illustrated for the case where $\theta=\pi/2$, which creates an approximate ellipse in $\hat{\mathbf{x}}'-\hat{\mathbf{y}}'$ plane, where semi-major and semi-minor axes are $a=A$  and $b=A\kappa h/2$,  respectively.}
   \label{fig:locomotion_analysis}
\end{figure}

Assuming that the robot's neutral surface generates a general one-dimensional wave motion, which may be expressed as 
\begin{equation}\label{general_wave_form}
\begin{aligned}
   y(x,t)=A\bigl(\cos (\kappa x)&\sin (\omega t +\theta) +\\ &\sin (\kappa x)\sin (\omega t)\bigr), 
\end{aligned}
\end{equation}
let us consider a fixed material point located at the lower boundary surface of the beam, which comes in contact with the floor. The point under consideration shares the same cross-sectional plane with a material point located at $x$ on the neutral surface.Thus their relative displacement is always perpendicular to the backbone curve $y(x,t)$ whose tangent is oriented by angle $\varphi(x,t)=\tan ^{-1} (\partial y/\partial x)$, as shown in Figure \ref{fig:locomotion_analysis}. In the robot's coordinate frame, $\hat{\mathbf{x}}'-\hat{\mathbf{y}}'$, we can describe the point's position as

\begin{equation}\label{point_position}
\begin{aligned}
\mathbf{r}_b'(x,t)=\left[x+\frac{h}{2}\sin\varphi\right] \hat{\mathbf{x}}'+\\
\left[y(x,t)-\frac{h}{2}\cos\varphi\right]\hat{\mathbf{y}}'. 
\end{aligned}
\end{equation}
In the general case, the robot alternates its contact points with the ground. Assuming kinematic constraint of no-slip contact, the absolute velocity of the instantaneous contact point,  in $\hat{\mathbf{x}}'$ direction, is zero. Thus, the robot's velocity $v_{f}(t)$ in $\hat{\mathbf{x}}'$ direction is equal in magnitude and opposite in direction to the contact point's instantaneous velocity $v_{c}^x(t)$, with respect to $\hat{\mathbf{x}}'-\hat{\mathbf{y}}'$ coordinate frame, so that $v_{f}(t)= -v_{c}^x(t)$  \cite{Zarrouk2016}. The expression for the robot's instantaneous velocity is given by

\begin{equation}\label{instantaneous_robot_velocity}
v_f(t)=\frac{A h \kappa  \omega  \sin (\theta ) }{2 \sqrt{\sin ^2( \omega t )+\sin^2(\omega t +\theta )}}.
\end{equation}
For the special case, where $\theta=\pi/2$, i.e. pure traveling wave, by assuming that $A^2\kappa^2\ll 1$, equation [\ref{point_position}] gives an approximate point's position which moves along the trajectory of an ellipse in $\hat{\mathbf{x}}'-\hat{\mathbf{y}}'$ plane with semi-major axis $a=A$ and semi-minor axis $b=A\kappa \frac{h}{2}$ (see Figure \ref{fig:locomotion_analysis}). The velocity in this case becomes constant, where $v_f(t)=v_f=A h \kappa  \omega/2$. Another two special cases are when $\theta=\{0,\pi\}$, meaning pure standing wave, in which case the velocity of the robot vanishes.

Next, we study the effect of the phase shift $\theta$ on the net propagation speed. For an arbitrary phase shift $\theta$ the robot's velocity $v_{f}(t)$ oscillates and is not constant, as stated in equation [\ref{instantaneous_robot_velocity}].  The average velocity of the robot is given by,

\begin{equation}\label{average_velocity}
\bar{v}_f=\frac{1}{T}\int_{0}^{T} v_f(t) \,dt.
\end{equation}
where $T=1/f$ is the time cycle. Figure \ref{fig:locomotion_analysis}(b) examines the effect of the temporal phase shift on the robot's velocity for a fixed $\hat{p}_1\approx\hat{p}_2$, under the assumptions of quasi-static motion and no-slip kinematic constraint, visualizing that the maximal velocity is attained for $\theta\approx\pi/2$, where the deformation pattern is closest to pure traveling wave as it was seen above, in Figure \ref{fig:IdentificationPreCurved}(d). Moreover, the velocity is approximately zero when $\theta=\{0, \pi\}$, which corresponds to a standing wave. An analytic result which was obtained in equation [\ref{average_velocity}] is overlaid in Figure \ref{fig:locomotion_analysis}(b), showing a good agreement.

For the special case, where $\theta=\pi/2$, i.e. pure traveling wave, Figure \ref{fig:locomotion_analysis}(c) depicts the trace of a point located at a distance of $h/2$ from the neutral surface, both experimentally and analytically according to

\begin{equation}\label{point_position_lab_frame}
\begin{aligned}
\mathbf{r}_b(x,t)\cdot \hat{\mathbf{x}}=&\frac{1}{2} A h \kappa  \omega t+\\&\frac{A\kappa h\sin(\omega t -\kappa x)}{2\sqrt{1+A^2 \kappa^2\sin^2(\omega t -\kappa x)}}. 
\end{aligned}
\end{equation}

\subsection*{Demonstrations}

\begin{figure*}
    \centering
       %trim={<left> <lower> <right> <upper>}

      %\includegraphics[clip, trim=0cm 20cm 0cm 0cm,width=1\linewidth]{Figure9_demonstration_unite.pdf}
    \includegraphics[width=1\linewidth]{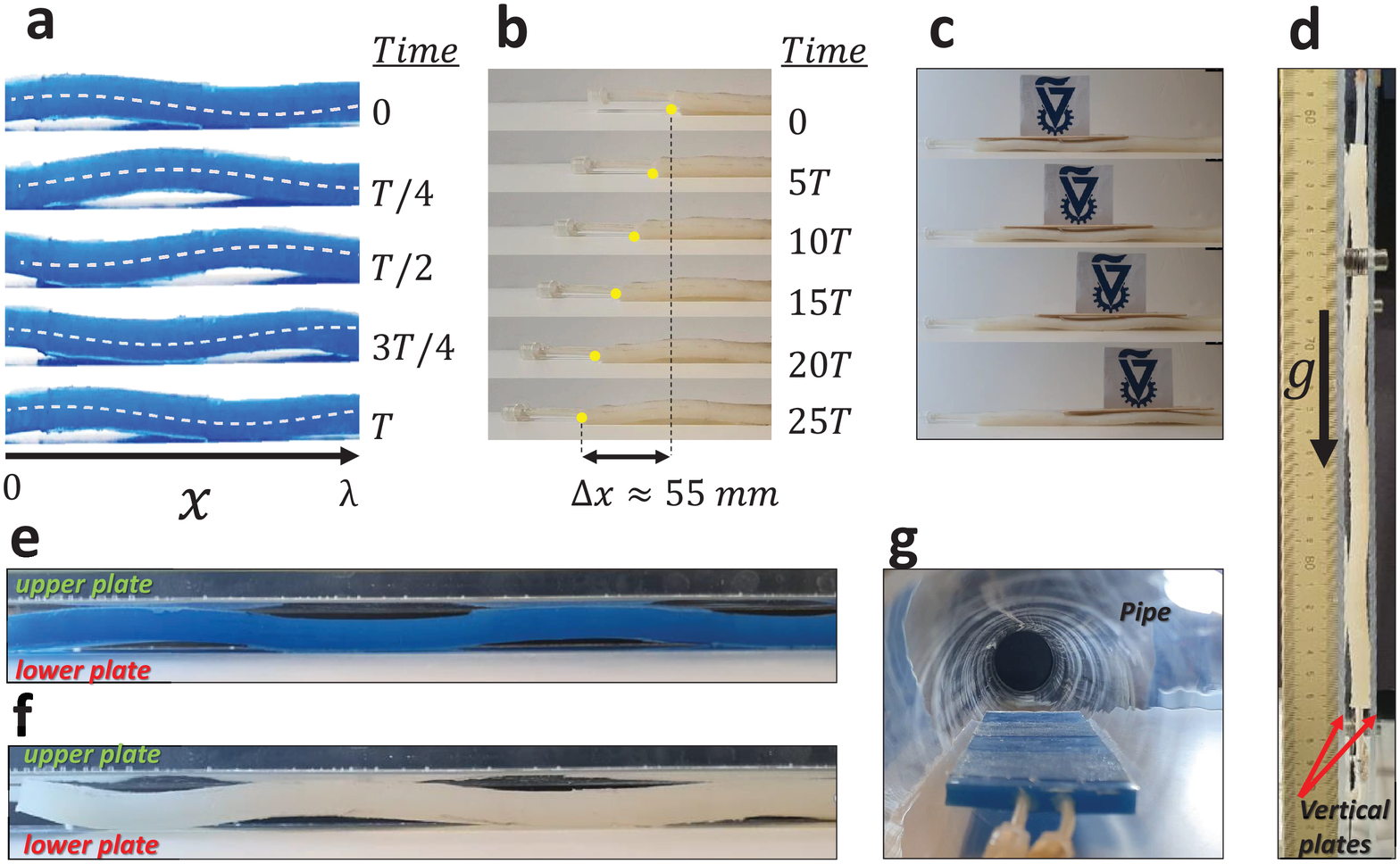}
    \caption[Robots' snapshots during motion.]{Robots' snapshots during motion. \textbf{(a)} Robot's shape during one time-cycle $T$, illustrated by the straight robot. Analytic wave-shape is marked by dashed white line. \textbf{(b)} Pre-curved robot propagation during 25 time cycles which moves $55$ mm with an average velocity of $v_{f}\approx 2.2$ mm/cycle. \textbf{(c)} The pre-curved robot is held at the right end, acting as a conveyor carrying a small block with the sign of the Technion logo in the opposite direction to the wave. \textbf{(d)} Pre-curved robot is climbing vertically between two plates. \textbf{(e)} Straight and \textbf{(f)} pre-curved robots propagating between two plates. \textbf{(g)} Straight robot crawling through a tube.}
   \label{fig:robot_snaps}
\end{figure*}

Visualization of the traveling wave motion is illustrated by the straight robot in Figure \ref{fig:robot_snaps}(a), where several snapshots were taken during one time-cycle at $t=0,T/4,T/2,3T/4,T$, where one can observe that the peak is advancing from left to right. Figure~\ref{fig:robot_snaps}(b) illustrates pre-curved robot's propagation on a horizontal surface, with an average velocity of $v_{f}\approx 2.2$ mm/cycle.

Several additional scenarios were tested in order to demonstrate the robot's locomotion capabilities. When one end is held fixed, the robot can act as a conveyor in order to transport objects in the direction opposing the wave's propagation.  Such configuration is illustrated in Figure \ref{fig:robot_snaps}(c), where the oscillating pre-curved robot is attached at the right end, and its wave direction is from right to left. The robot is conveying a small block with the sign of the Technion logo in the opposite direction to the wave, i.e. from left to right.
An additional capability of traveling wave soft robots is to exploit their flexibility in order to move through narrow spaces. Soft robot climbing vertically (see Figure \ref{fig:robot_snaps}(d)) and crawling horizontally (see Figure \ref{fig:robot_snaps}(e)-(f)) in a narrow gap between two parallel plates are demonstrated. In addition, the traveling wave soft robots' flexibility and compliance can also be utilized for crawling in a pipe, as shown in Figure \ref{fig:robot_snaps}(g). A supplementary movie includes illustrations of the described scenarios.

%TC:ignore
\section{Conclusion}

The formation of traveling waves in slender elastic configurations is of applicative importance and may pave the way to progress in many fields that utilize propulsion in unpredictable and narrow passages such as disaster areas or within vascular systems. Moving waves allow for thrust to be applied throughout the robot, thus to self-propel in a long and winding path.

 In this work, we present a model-based method to design and generate continuous traveling waves in elastic structures with a fluidic network using only two controlled pressure inlets. This reduction in inlets simplifies the implementation in confined spaces and on-board platforms due to fewer hardware components and overall weight. First, we provided theoretical analysis that guided the design of the channel network distribution and the corresponding continuous pressure inlets. Then, using standard and simple casting techniques, we fabricated two types of soft robots capable of generating bi-directional traveling waves. Next, we conducted experiments, analyzed the generated waves, and applied a method of inlet calibration. Finally, we provided a kinematic analysis that allows gait planning and demonstrated robot locomotion in various scenarios. The experimental results agree well with the simplified theoretical models. However, deviations of experimental values from the theoretical model are associated with design compromises, fabrication inaccuracies, and unmodeled effects such as inertia, external forces, contacts, and non-linearity of the elastic structure that future works may address.

%%%%%%%%%%%%%%%%%%%%%%%%%%%%%%%%%%%%%%%%%%%%%%%%%%%%%%%%%%%%%%%%%%%%%%%%%%

\section*{Acknowledgements} 
All authors contributed to the design and implementation of the research, to the analysis of the results and to the writing of the manuscript.

\section*{Author Disclosure Statement}
No competing financial interests exist.

%%%%%%%%%%%%%%%%%%%%%%%%%%%%%%%%%%%%%%%%%%%%%%%%%%%%%%%%%%%%%%%%%%%%%%%%%%

\bibliographystyle{ieeetr}
\bibliography{pnas_byb.bib}

\section*{Supplementary Information}

\subsection*{Fabrication and Experimental setup}

The soft robot fabrication process is presented in Figure \ref{fig:FabExp}. According to theoretical model guidelines, two sets of 3D printed core (PVA, BCN Sigma) in a serpentine geometry are embedded into an elastomer cast (DragonSkin20, Smooth-On Inc. See Figure \ref{fig:FabExp}(a)) and positioned at the center of the structures' cross-section (See Figure \ref{fig:FabExp}(b)). Embedded channels properties are: length $l_c=1600$ mm, and radius $r_c=1$ mm. The elastic robot has rectangular cross-section with physical properties: length $l_s=280$ mm, cross-section: height $h_s=9$ mm, width $b_s=70$ mm, Young's Modulus $E=0.97$ MPa, and density $\rho =1080$ kg/m$^3$. The PVA core is then dissolved by warm water ($50^{\circ}$C) using a circulating water pump {fig:FabExp}(d).

Attaching limited-strain layers to the outer faces ( See Figure {fig:FabExp}(e)) moves the neutral surface from the center to the outer faces thus creating thinner structures. For this reason,  limited-strain layers are glued with the same cast material on the elastic structure, alternately at top and bottom faces in order to determine a new neutral surface and curvature direction, as illustrated in dashed lines in Figure \ref{fig:Design}. The width of these layers is the same as the robot's width, and their length, $l_f$, depends on the curvature and can be calculated by using \begin{equation}\label{FiberLength}
   l_f=\int_{0}^{\lambda/2} \sqrt{1+\left(\frac{d y_0}{d x}\right)^2} dx,
\end{equation}
where $y_0(x)$ is the robot's pre-shape as stated in equation [\ref{PreCurvedShape}]. Note that for the straight robot case, namely when $y_0(x)=0$,  each limited-strain fiber length is $\lambda/2$. Silicone tubes were glued using silicone rubber adhesive (Sil-Poxy, Smooth-On Inc.), and Luer connectors were attached to enable convenient inlet connections. Robot's top-view displaying two sets of channels and limited-strain layers is shown in Figure \ref{fig:Design}(d).

Figure \ref{fig:FabExp}(g) is illustrating the experimental setup. The robot was placed on a flat surface, and a pressure controller (Elveflow OB1) was connected to its inlets. A camera (Logitech C920, HD, 30fps ) was placed perpendicular to the deflection plane and captured frames simultaneously to actuation. In order to extract deformation signal, we used MATLAB\textsuperscript{\textregistered} image processing toolbox.

\subsection*{Deflection calibration under negative gauge pressure}

\begin{figure}[h]
    \centering
   %trim={<left> <lower> <right> <upper>}
    %\includegraphics[clip, trim=0cm 2.5cm 6cm 0cm,width=1\linewidth]{figures/FigDeformations2.pdf}
    %\includegraphics[width=1\linewidth]{figures/ResultsActuator.eps}
    \includegraphics[width=1\linewidth]{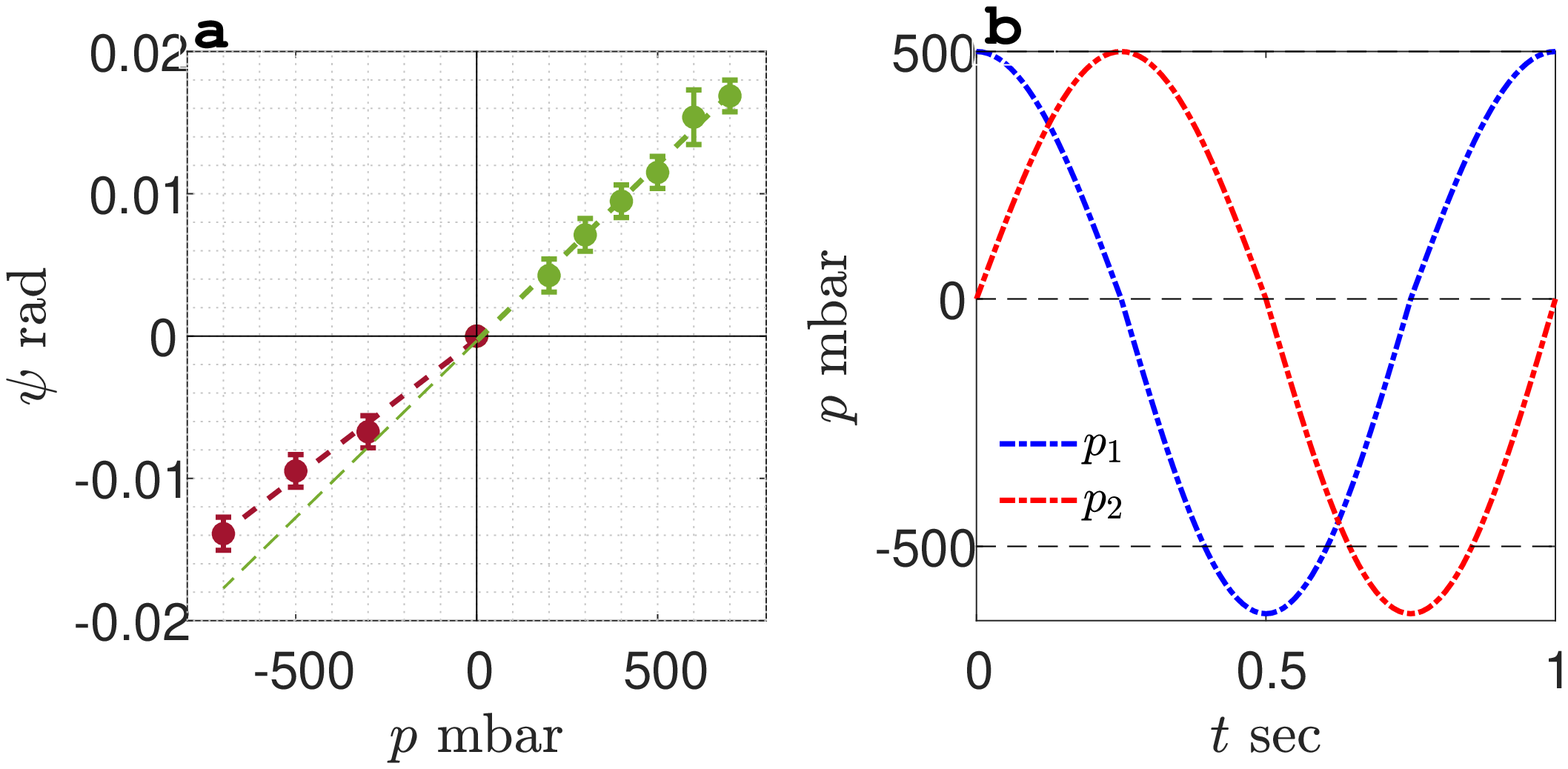}
    \caption[SI Figure-Deflection calibration.]{SI Figure- Deflection calibration. \textbf{(a)} Deflection angle, $\Psi$, versus pressure, $p$, for positive (green markers) and negative (bordeaux markers) ranges. Dashed lines represent slopes obtained by linear regression ($R^2=0.993$ - positive range, $R^2=0.989$ - negative range). The results present a greater slope for the positive range than for the negative range, by a factor of $\approx 1.27$. \textbf{(b)} Adjusted inlet pressures, $p_i(t)$, $i=1,2$, according to equation [\ref{PressureCompensation}].  Amplitudes for negative gauge pressure are higher than for positive gauge pressure in order to compensate for the lower deflection. }
   \label{fig:NPslopes}
\end{figure}

This section describes a calibration process that carried out to compensate for smaller deflections due to negative gauge pressures.
Applying pressure in the elastic structure inlets causes deformation. The change in beam slope due to a single channel is defined by $\psi$. The gradient $\gamma_+\triangleq \partial \psi/\partial (p/E)_+$ is assumed to be approximately constant for a certain positive range of pressures \cite{Matia2015}. However, it was noticed that for the negative gauge pressure the range of the gradient $\gamma_-$  has a moderately, but noticeable, smaller slope. In order to validate this assumption and to obtain the ratio between the slopes we performed a calibration experiment as follows: the beam was clamped and placed, where the deformation plane was perpendicular to gravity. Inlet pressures of $-700,-500,-300,200,300,400,500,600$ and $700$ mbar were applied and deformation was measured at the end of a section consisting from 12 channels.  Assuming that the contribution of each channel is equal, the total angle change was calculated and divided by 12 to obtain the average contribution of a single channel. Slopes were obtained using linear regression with  $R^2=0.993$ for the positive range, and $R^2=0.989$ for the negative range. The results present a greater slope for the positive range than for the negative range by a factor of $\approx 1.27$. We statically pressurized the robot by a negative gauge pressure with amplitude greater by 1.27 than for a positive gauge pressure and measured the static deformation amplitude, to validate the calibration results. Calibration results appear in Figure \ref{fig:NPslopes}.

\subsection*{Curve fitting and TWR}

This section describes in detail the method used to analyze wave signals by tempo-spatial decomposition and quantify the propagation via traveling wave ratio (TWR). This practice is well established and demonstrated in \cite{VERED2021107515, bucher2019experimental, bucher2004estimating,setter2014propulsion}.

\subsubsection*{Curve fitting}
The varying deformation signal can be decomposed in time and space. We assume the deformation takes the form of a truncated Fourier series in $x$ as

\begin{equation}\label{Fitting_series_appendix}
    y(x,t)=d(x)+\sum_{n=1}^{N_x}y_n(x,t)+r(x,t),
\end{equation}
where $d(x)$ accounts for a spatial shape, $y_n(x,t)$ is the oscillating function of the $n^{th}$ spatial mode, and $r(x,t)$ is the residual function. The function $y_n(x,t)$  can be decomposed to time-space domains as
\begin{equation}\label{Fitting_one_harmony}
    y_n(x,t)=a_n(t)\cos(\kappa_n x)+b_n(t)\sin(\kappa_n x),
\end{equation}
where $\kappa=2\pi/\lambda$  and $\kappa_n=n\kappa, \forall  n=1,2,...,N_x$. Each spatial harmony can be decomposed such as
\footnotesize
\begin{equation}\label{atbt_appendix}
\begin{aligned}
a_n=\sum_{m=1}^{N_t} \left( \tilde{A}_{m}^n\cos{(\omega_m t)}+\tilde{B}_{m}^n\sin{(\omega_m t)}\right)+r_n^a(t),\\
b_n=\sum_{m=1}^{N_t}\left(\tilde{C}_{m}^n\cos{(\omega_m t)}+\tilde{D}_{m}^n\sin{(\omega_m t)}\right)+r_n^b(t),\\
\end{aligned}
\end{equation}
\normalsize
where $\omega=2\pi f$  and  $\omega_m =m\omega,\forall m=1,2,...,N_t$ . A spatial function $d(x)$ is added to account for robot's curvature shape
\begin{equation}\label{deviationx_appendix}
  \begin{gathered}
d(x)=\sum_{n=1}^{N_x}\left(d_n^c\cos(\kappa _n x)+d_n^s\sin(\kappa _n x)\right) \approx\\ \approx D\sin\left(\kappa _1 x+\vartheta\right)\\
D=\sqrt{{d_1^c}^2+{d_1^s}^2}  \quad\text{,}\quad \vartheta= tg^{-1}\left(d_1^c/d_1\right)\\
  \end{gathered}
\end{equation}
and may be approximated to the designed pre-determined curvature associated with the first wave-number. The initially straight robot oscillates about the horizontal axis, i.e. when $D=0$. In the case of the pre-curved robot due to its initial curvature, $D$ depends on the applied pressures. When applied pressures match working point pressures, the pre-curved robot also oscillates about the horizontal axis as desired.

To obtain the scalars $\tilde{A}_m^n,\tilde{B}_m^n,\tilde{C}_m^n,\tilde{D}_m^n$ and $D$, we first solve a linear least square problem as stated in equation [\ref{LeastSquare_x_appendix}] below to obtain $a_n(t)$ and $b_n(t)$. The linear least square problem was solved for a range of wave-numbers $\kappa$, and ultimately determined in a sense that minimizes the residual.

\begin{equation}\label{LeastSquare_x_appendix}
\begin{aligned}
  \mathbf{a}=\mathbf{A}^\dagger\mathbf{y}=(\mathbf{A}^T\mathbf{A})^{-1}\mathbf{A}^T\mathbf{y},\\
  \end{aligned}
\end{equation}
where the coefficients vector is

\begin{equation}\label{a_vec_coefficients_appendix}
\mathbf{a}=\begin{bmatrix}
 a_1(t) & b_1(t) & ... & a_{N_x}(t) & b_{N_x}(t) \\
\end{bmatrix}
\end{equation}
and
\footnotesize
\begin{equation}\label{A_model_matrix_appendix}
\mathbf{A}=\begin{bmatrix}
 \cos (\kappa x) &\sin (\kappa x) & ... & \cos (N_x\kappa x) & \sin ( N_x\kappa x) \\
\end{bmatrix}
\end{equation}
\normalsize
is the spatial model matrix.  To estimate $a_n(t)$ and $b_n(t)$ we then solved a similar least square problem as stated in equation [\ref{LeastSquare_t_appendix}] for a range of angular frequency $\omega$ that determined where residuals $r_n^a(t)$ and $r_n^b(t)$ are minimized
\begin{equation}\label{LeastSquare_t_appendix}
\begin{aligned}
  \mathbf{q}_n^a=\mathbf{Q}^\dagger\mathbf{a}_n(t), && \mathbf{q}_n^b=\mathbf{Q}^\dagger\mathbf{b}_n(t),\\
  \end{aligned}
\end{equation}
where the coefficient vectors are

\begin{equation}\label{wave_scalar_coefficients_appendix}
\begin{aligned}
\mathbf{q}_n^a=\begin{bmatrix}
 \tilde{A}_{1}^n & \tilde{B}_{1}^n  & ... &\tilde{A}_{N_t}^n & \tilde{B}_{N_t}^n &d_n^c \\
\end{bmatrix}\\
\mathbf{q}_n^b=\begin{bmatrix}
 \tilde{C}_{1}^n & \tilde{D}_{1}^n  & ... &\tilde{C}_{N_t}^n & \tilde{D}_{N_t}^n &d_n^s \\
\end{bmatrix}\\
\end{aligned}
\end{equation}and 
\footnotesize
\begin{equation}\label{Q_model_matrix_appendix}
\mathbf{Q}=\begin{bmatrix}
\cos (\omega t) &\sin (\omega t) & ... & \cos (N_t\omega t) & \sin ( N_t\omega t)& 1  \\
\end{bmatrix}
\end{equation}
\normalsize
is the temporal model matrix.

Using the tempo-spatial decomposition we were able to fit an analytic function to the robot's oscillating signal using $N_x=4$ and $N_t=3$, where higher wave-numbers and wave frequencies had very small amplitudes and minor contribution to the fitted signal. Moreover, the single wave associated with the designed wave-number and the excitation frequency was greater than the other modes by an order of magnitude, hence we are interested only in this single wave which is further analyzed.

\subsubsection*{TWR}

The traveling wave ratio (TWR) is a scalar parameter that quantifies a single wave propagation characteristics. The modal decomposition allows analyzing a single wave-number and a single frequency wave in order to calculate its TWR. One dimensional wave can be written as a combination of two traveling waves propagating in opposite directions
\begin{equation}\label{complex-amplitudes_SI}
    y(x,t)=\Re\left(U_+e^{i (\omega t-\kappa x)}+U_-e^{i (\omega t+\kappa x)}\right),
\end{equation}
where

\begin{equation}\label{complex_amplitudes_scalars}
   \begin{aligned}
    U_+=U_{+}^R+iU_{+}^I,\\      
    U_-=U_{-}^R+iU_{-}^I
  \end{aligned}
\end{equation}
are the complex forward and backward wave's amplitudes, and $U_{+}^R, U_{+}^I, U_{-}^R, U_{-}^I$ are scalars. We can define the traveling wave ratio (TWR) as the ratio between the standing wave and the mixed wave amplitude

\begin{equation}\label{TWR_definition_SI}
    TWR=1-|SWR^{-1}|=1-\frac{||U_+|-|U_-||}{|U_+|+|U_-|}.
\end{equation}
This bounded ratio $0 \leq TWR \leq 1$, defines wave's propagation where $TWR=0$ denotes pure traveling wave, whereas $TWR=1$ denotes pure standing wave. Any other value between $0$ to $1$ expresses a mixed traveling and standing wave. Moreover, the wave's direction can be obtained by the sign of $|U_+|-|U_-|$, positive or negative corresponds to forward or backward direction, respectively.  To calculate TWR as presented in equation [\ref{TWR_definition_SI}], the complex-amplitudes of the forward and the backward waves can be obtained using the coefficients obtained in equation [\ref{wave_scalar_coefficients_appendix}] and the following transformation, 

\begin{equation}\label{coefficient_transformation}
    \begin{bmatrix}
U_{+}^R & U_{+}^I \\
U_{-}^R & U_{-}^I 
\end{bmatrix}=\frac{1}{2}\begin{bmatrix}
1 & 1 \\
1 & -1 
\end{bmatrix}\begin{bmatrix}
\tilde{A}_m^n & -\tilde{B}_m^n \\
\tilde{D}_m^n & \tilde{C}_m^n 
\end{bmatrix}
\end{equation}
combined with equation [\ref{complex_amplitudes_scalars}].

\subsection*{Locomotion}

This section provides complementary and expanded equations for "Traveling wave-robot locomotion" chapter.

\subsubsection*{Mixed-wave: Arbitrary temporal-phase-shift $\theta$}

The general case for an arbitrary temporal phase shift is discussed. Assuming that the robot's neutral surface generates a general one-dimensional wave motion which may be expressed as,

\begin{equation}\label{general_wave_form_appendix}
\begin{aligned}
   y(x,t)=A\bigl(\cos (\kappa x)\sin (\omega t +\theta) +\\ \sin (\kappa x)\sin (\omega t)\bigr),
\end{aligned}
\end{equation}
we may consider a fixed material point located at the lower boundary surface of the beam, which comes in contact with the floor. The point under consideration shares the same cross-sectional plane with a material point which is located at $x$ on the neutral surface.Thus their relative displacement is always perpendicular to the backbone curve $y(x,t)$ whose tangent is oriented by angle $\varphi(x,t)=\tan ^{-1} (\partial y/\partial x)$, as shown in Figure \ref{fig:locomotion_analysis}. In the robot's coordinate frame we can describe the point's position as

\begin{equation}\label{point_position_appendix}
\begin{aligned}
\mathbf{r}_b'(x,t)=\left[x+\frac{h}{2}\sin\varphi\right] \hat{\mathbf{x}}'+\\\left[y(x,t)-\frac{h}{2}\cos\varphi\right]\hat{\mathbf{y}}'. 
\end{aligned}
\end{equation}
The positions of the robot's lower boundary surface points are given by
\footnotesize
\begin{multline}\label{boundary_point_position_full_x_direction}
r_{b}^x(x,t)=\mathbf{r}_b' \cdot \hat{\mathbf{x}}'=x+\\
\frac{0.5A h (\kappa  \sin (\omega t ) \cos (\kappa  x)-\kappa  \sin (\kappa  x) \sin (\omega t + \theta
   ))}{\sqrt{(A\kappa)^2 ( \sin (\omega t ) \cos (\kappa  x)-  \sin (\kappa  x) \sin (\omega t +\theta ))^2+1}}.
\end{multline}
\normalsize
Temporal differentiation of equation [\ref{boundary_point_position_full_x_direction}] yields the velocity of the lower boundary surface points, relative to robot's coordinate system as,
\footnotesize
\begin{multline}\label{boundary_point_velocity_full_appendix}
v_{b}^x(x,t)=\mathbf{v}_b'(x,t) \cdot \hat{\mathbf{x}}'=\frac{d\mathbf{r}_{b}'}{dt}\cdot \hat{\mathbf{x}}' =\\
\frac{0.5A h \kappa  \omega  (\cos (\omega t ) \cos (\kappa  x)-\sin (\kappa  x) \cos (\omega t + \theta
   ))}{\left((A\kappa)^2 ( \sin (\omega t ) \cos (\kappa  x)- \sin (\kappa  x) \sin (\omega t +\theta ))^2+1\right)^{3/2}}.
\end{multline}
\normalsize
The time-varying material point on the lower boundary surface which comes in contact with the ground is located at $x_c(t)$, where the height is minimal and $\partial y/\partial x=0$. We may find $x_c(t)$ by employing the following expression,

\begin{equation}\label{instantaneous_contact_point_location_appendix}
x_c(t)=\frac{1}{\kappa}\tan^{-1}\left(\frac{\sin(\omega t)}{\sin(\omega t+\theta)}\right).
\end{equation}
Substituting equation [\ref{instantaneous_contact_point_location_appendix}] into equation [\ref{boundary_point_velocity_full_appendix}] yields the instantaneous velocity of the contact point $v_c^x(t)$, with respect to robot's coordinate system,

\begin{equation}\label{instantaneous_contact_point_velocity_appendix}
v_c^x(t)=-\frac{A h \kappa  \omega  \sin (\theta ) }{2 \sqrt{\sin ^2( \omega t )+\sin^2(\omega t +\theta )}}.
\end{equation}
Assuming the kinematic constraint of no-slip contact, the absolute velocity of the contact point, located at $x_c(t)$ is zero. Thus, the robot's velocity in $\hat{\mathbf{x}}'$ direction $v_f(t)$ is equal in magnitude and opposite in direction to the contact point's velocity, so that $v_{f}(t)= -v_{c}^x(t)$  \cite{Zarrouk2016}.

Moreover, the average velocity of the robot is given by,

\begin{equation}\label{average_velocity_appendix}
\bar{v}_f=\frac{1}{T}\int_{0}^{T} v_f(t) \,dt,
\end{equation}
where $T$ is the time cycle.

\subsubsection*{Pure traveling-wave: Temporal-phase-shift $\theta=\pi/2$}

For the special case where $\theta=\pi/2$, i.e. pure traveling wave, equation [\ref{point_position_appendix}] takes the following form,
\footnotesize
\begin{multline}\label{point_position_pureTW_appendix}
\mathbf{r}_b' (x,t,\theta=\pi/2)=\\ \left[x+\frac{A\kappa h\sin(\omega t -\kappa x)}{2\sqrt{1+A^2 \kappa^2\sin^2(\omega t -\kappa x)}}\right] \hat{\mathbf{x}}'+\\\left[A\cos(\omega t-\kappa x)-\frac{h}{2\sqrt{1+A^2 \kappa^2\sin^2(\omega t -\kappa x)}} \right]\hat{\mathbf{y}}'.
\end{multline}
\normalsize
Assuming that $A^2\kappa^2\ll 1$, we get from equation [\ref{point_position_pureTW_appendix}] that

\begin{equation}\label{point_position_pureTW_approx_appendix}
\begin{aligned}
\mathbf{r}_b' (x,t,\theta=\pi/2) \approx \\\left[x+\frac{h}{2}A\kappa\sin(\omega t -\kappa x)\right] \hat{\mathbf{x}}'+\\\left[A\cos(\omega t-\kappa x)-\frac{h}{2} \right]\hat{\mathbf{y}}',
\end{aligned}
\end{equation}
which gives an approximate point's position which moves along the trajectory of an ellipse in $\hat{\mathbf{x}}'-\hat{\mathbf{y}}'$ plane with semi-major axis $a=A$ and semi-minor axis  $b=A\kappa \frac{h}{2}$, as illustrated in Figure \ref{fig:locomotion_analysis}.

The velocity of the boundary point, with respect to robot's coordinates system, for the pure traveling wave case can be obtained either by substituting $\theta=\pi/2$ into equation [\ref{boundary_point_velocity_full_appendix}] or by differentiating equation [\ref{point_position_pureTW_appendix}] with respect to time. In any case, we get that
\footnotesize
\begin{multline}\label{point_velocity_pureTW_appendix}
\mathbf{v}_b'(x,t,\theta=\pi/2) = \\\frac{Ah\kappa \omega \cos(\omega t -\kappa x)}{2(1+A^2\kappa ^2 \sin^2(\omega t-\kappa x))^{3/2}}\hat{\mathbf{x}}'
-A\omega\sin(\omega t-\kappa x)\cdot\\\cdot\left[1-\frac{A\kappa ^2 h\cos(\omega t-\kappa x) }{2(1+A^2\kappa ^2 \sin^2(\omega t-\kappa x))^{3/2}}\right]\hat{\mathbf{y}}'.
\end{multline}
\normalsize
 In order to obtain the instantaneous contact point velocity, $v_{c}^x$, with respect to robot's coordinates system, one can substitute $\theta=\pi/2$ into equation [\ref{instantaneous_contact_point_velocity_appendix}] or to substitute equation [\ref{instantaneous_contact_point_location_appendix}] into equation [\ref{point_velocity_pureTW_appendix}] which

yields
\begin{equation}\label{v_c_TW_appendix}
v_{c}^x=-A\kappa\omega h/2,
\end{equation}
where in this case, the velocity of the contact point is constant and equal in magnitude, but opposite in direction, to the robot's constant velocity, namely

\begin{equation}\label{v_f_TW_appendix}
\bar{v}_f=v_{f}= -v_{c}^x=\frac{1}{2}A\kappa\omega h,
\end{equation}
where $\bar{v}_f$ is the robot's average velocity.

%TC:endignore
\end{multicols}
\end{document}